\documentclass[12pt]{iopart}

\usepackage[pdftex]{graphicx}

\usepackage{iopams}

\usepackage[square,numbers,sort&compress]{natbib}

\begin{document}

\title[J Ehrig, E P Petrov and P Schwille]{Phase separation and near-critical fluctuations in two-component lipid membranes: Monte Carlo simulations on experimentally relevant scales}

\author{Jens Ehrig, Eugene P Petrov and Petra Schwille}

\address{Biophysics, BIOTEC, Technische Universit\"{a}t Dresden, Tatzberg 47/49, 01307 Dresden, Germany}
\ead{petrov@biotec.tu-dresden.de}

\begin{abstract}
By means of lattice-based Monte Carlo simulations, we address properties of two-component lipid membranes on the experimentally relevant spatial scales of order of a micrometer and time intervals of order of a second, using DMPC/DSPC lipid mixtures as a model system. Our large-scale simulations allowed us to obtain important results previously not reported in simulation studies of lipid membranes. We find that, within a certain range of lipid compositions, the phase transition from the fluid phase to the fluid--gel phase coexistence proceeds via near-critical fluctuations, while for other lipid compositions this phase transition has a quasi-abrupt character. In the presence of near-critical fluctuations, transient subdiffusion of lipid molecules is observed. These features of the system are stable with respect to perturbations in lipid interaction parameters used in our simulations. The line tension characterizing lipid domains in the fluid--gel coexistence region is found to be in the pN range. When approaching the critical point, the line tension, the inverse correlation length of fluid--gel spatial fluctuations, and the corresponding inverse order parameter susceptibility of the membrane vanish. All these results are in agreement with recent experimental findings for model lipid membranes. Our analysis of the domain coarsening dynamics after an abrupt quench of the membrane to the fluid--gel coexistence region reveals that lateral diffusion of lipids plays an important role in the fluid--gel phase separation process.
\end{abstract}


\vspace{2pc}

\vspace{2pc}
\noindent{\textit{New Journal of Physics}, 2011, accepted for publication}

\maketitle

\section{Introduction}
The membrane plays a very important role in the cell, not only by defining its boundaries and boundaries of the cell organelles, but also by taking an active part in cell functioning, as many key processes take place in or across the cell membrane. The plasma membrane is a very complex system comprised of lipids, polysaccharides, and proteins, all strongly interacting with each other. These lipid--lipid and lipid--protein interactions induce the lateral microheterogeneity of the cell membrane, which is believed to be crucial in providing its high functionality \cite{Vereb2003}. The idea of the function-related lateral microheterogeneity of the membrane was formulated by Simons and Ikonen \cite{Simons1997} in the form of the concept of lipid rafts. The original concept of lipid rafts involved the dynamic clustering of sphingolipids and cholesterol, supposedly forming moving platforms in the cell membrane to which certain proteins would attach \cite{Simons1997}. Presently, the understanding is that the interaction of lipids with membrane proteins and the cytoskeleton, as well as effects of local curvature, play an important role in organization of the lateral microheterogeneity of the cell membrane \cite{Garg2009, Liu2006, Roux2005, Sorre2009}. Consequently, the term membrane rafts appears to be more appropriate \cite{Pike2006}.

In spite of the importance of lipid--protein interactions, the behaviour of the lipid component of the cell membrane is still believed to govern its properties to a large extent. In agreement with that, it was found that studies of model membrane systems -- both \textit{in vitro} \cite{Weissig2009} and \textit{in silico}, i.e., via numerical simulations \cite{Deserno2009, Elson2010, Heimburg2007, Mueller2006} -- give valuable information on the lateral organization and dynamics of lipid molecules, which provides a deeper insight into the structure and function of plasma membranes.

Numerical simulations are an extremely fruitful approach to membrane studies, giving access to information which is either difficult or even impossible to access using present-day experimental techniques. Depending on particular aims and goals, a wide arsenal of simulation techniques has been developed to study model membrane systems. A particular choice of the simulation approach largely depends on the amount of the fine molecular details, as well as spatial scales and time ranges of relevance for a particular problem addressed in a study. Broadly, the approaches to numerical simulations of lipid membranes include molecular dynamics methods and their coarse-grained versions, mean-field based continuum model simulations, and lattice-based methods.

Molecular dynamics simulations provide the most detailed approach to describe the structure and dynamics of membranes. They employ an atomistic description of the lipid membrane and thus, with the presently available computational resources, are able to cover spatial scales up to a few nm and time scales of $\leq 1$ $\mu$s \cite{Marrink2005, Shi2005, Niemelae2007, Bjelkmar2009}. By using coarse-grained molecular dynamics simulations which sacrifice the very fine atomistic details for the sake of a better computational efficiency \cite{Deserno2009, Mueller2006, Pandit2009} these scales can be extended to $\approx 100$ nm and $\approx 10$ $\mu$s. Further coarse-graining \cite{Yuan2010, Pasqua2010}, by representing a cluster of a few ten lipid molecules as one particle, allows for simulations of membranes consisting of $\sim 6000$ coarse-grained particles, which is equivalent to the total number of $\sim 60000$ lipid molecules in a simulated membrane. On the other hand, this type of coarse-graining does not allow one to address, e.g., diffusional motion of individual lipid molecules in the membrane.

Mean field-based continuum simulations \cite{Lowengrub2009, Wang2008}, based on solving evolution equations for a compositional or phase field, are primarily concerned with the issue of the phase separation behaviour of lipid membranes \cite{Reigada2008, Haataja2009, Fan2010a}. These simulations address much larger spatial scales of order of a few hundred nm. On the other hand, this approach does not seem to be able to reproduce the complete phase diagram of a real lipid mixture. Additionally, these simulations do not provide information on lateral diffusion in the membrane.

Continuum simulations of the lipid membrane organization can be made more realistic by using lipid interaction parameters and lipid chain order parameter libraries extracted from molecular dynamics simulations \cite{Tumaneng2010}. For ternary lipid membrane systems, these simulations can currently address spatial scales of a few tens of nm \cite{Tumaneng2010}. In view of the continuum character of the model, they do not, however, provide information on the lateral diffusion of membrane components.

Lattice-based Monte Carlo simulations constitute the computationally least demanding way to numerical modelling of lipid membranes. Not surprisingly, the first lattice-based simulations of the single-component lipid membrane date back to the early l980s \cite{Mouritsen1983}. Later, this approach based on either the ten-state Pink model or Ising-like models was successfully used to study phase transitions in one- and two-component lipid membranes, as well as the effects of lipid--protein and lipid--drug interactions on the phase separation in lipid membranes (see, e.g., \cite{Joergensen1993, Zhang1993, Sugar1994, Joergensen1995, Sugar1999, Hac2005}). Although these lattice-based approaches do not even attempt at reproducing the membrane structure and dynamics in full detail on the atomic and molecular scale, they provide a surprisingly realistic description of the phase separation and dynamics in lipid membranes. Additionally, because of the simplified description of the membrane, lattice-based simulations are capable of describing the membrane behaviour on much larger spatial scales and substantially longer time intervals than the above-mentioned atomistic and coarse-grained simulations with the same computational expenses. While the previously published lattice-based simulations addressed one- or two-component lipid membranes on spatial scales from ten to hundred nm, the recent rapid progress in computer hardware presently allows one to substantially expand the spatial scales and time intervals addressed in lattice-based simulations.
	
The aim of the present study is to address the properties of model lipid membranes using a reasonably simple, but still realistic enough, model on the experimentally relevant spatial scales of order of a micrometer and time scales of order of a second. Simulations on these spatial scales and time intervals should provide results which can be directly compared with experimental results obtained using optical microscopy-based methods routinely used in membrane studies. Our goal was, therefore, to develop a lattice-based Monte Carlo simulation approach that would not only provide an adequate description of the membrane phase behaviour and membrane dynamics, but in addition would be computationally efficient in order to reach the required spatial and temporal scales.

\begin{figure}[!t]
\begin{center}
\includegraphics{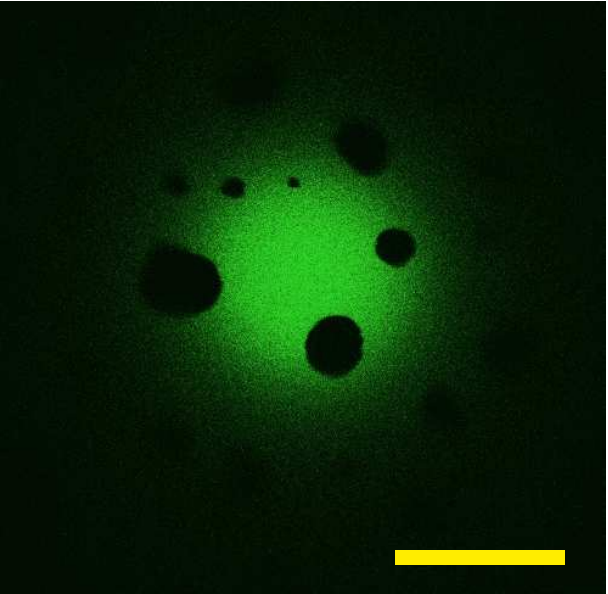}
\caption{Confocal fluorescence microscopy image of the upper pole of a giant unilamellar vesicle (DMPC/DSPC 50/50) exhibiting fluid--gel phase separation. Gel-phase domains, which coarsen and coalesce with time, appear on the image as dark areas. Scale bar: 20 $\mu$m.}
\label{Fig_GUV}
\end{center}
\end{figure}

We chose to study membranes consisting of a binary mixture of DMPC (1,2-dimyristoyl-\textit{sn}-glycero-3-phosphocholine, $T_{\rm m} = 297$ K) and DSPC (1,2-distearoyl-\textit{sn}-glycero-3-phosphocholine, $T_{\rm m} = 328$ K), which for a range of lipid compositions and temperatures shows coexistence of fluid and gel phases (figure \ref{Fig_GUV}). The phase diagram of this system has been extensively studied experimentally (see, e.g., \cite{Sugar1999, Hac2005, Shimshick1973, Mabrey1976, Lentz1976, Dijck1977, Wilkinson1979, Nibu1995, Sugar2000}). Some aspects of the phase behaviour of the same lipid mixture have been previously addressed in a series of lattice-based Monte Carlo simulations (see, e.g., \cite{Sugar1999, Hac2005, Sugar2000}). Therefore, the choice of the well-studied DMPC/DSPC lipid systems as an object of our study allowed for a direct comparison of the results of our numerical simulations with experimental data, as well as with results obtained in simulation studies of the same lipid system carried out by other groups.

In our recent work, using the same MC simulation approach as in the present paper, we addressed the aspects of lateral diffusion in the membrane with a focus on the effects of near-critical fluctuations in lipid membranes and the effects of cytoskeleton \cite{Ehrig2011}.

In the present paper, the main emphasis is on the phase diagram of the DMPC/DSPC lipid system, including the near-critical behaviour of the membrane close to its critical point, as well as the properties of membrane domains, including the line tension of the phase interface, and domain growth kinetics.

\section{Materials and Methods}

\subsection{Experimental}

The lipids 1,2-dimyristoyl-\textit{sn}-glycero-3-phospho\-choline (DMPC), 1,2-distearoyl-\textit{sn}-glycero-3-phospho\-choline (DSPC) and the fluorescent lipid marker 1,2-dipalmitoyl-\textit{sn}-glycero-3-phospho\-ethanol\-amine-N-(lissamine rhodamine B sulfonyl) (N-Rhod-DPPE) were purchased from Avanti Polar Lipids (Alabaster, AL), diluted in a chloroform-methanol (2:1) mixture, aliquoted, and stored at $-20^{\circ}{\rm C}$ in argon atmosphere. Giant unilamellar vesicles (GUVs) were prepared by electroformation on a platinum wire \cite{Angelova1986} and were used for visualization of phase separation in two-component DMPC/DSPC lipid membranes by means of laser scanning fluorescence microscopy. Multilamellar vesicle suspensions were obtained from rehydration of a dry lipid film. Differential scanning calorimetry (DSC) experiments were performed on suspensions of DMPC/DSPC multilamellar vesicles using a VP-DSC calorimeter (MicroCal, Northampton, MA) with a scan rate of $2-3$ K/h and a medium (``mid'') feedback.

The temperatures of the onset and completion of the phase transition were determined from experimentally measured excess heat capacity curves using an empirical approach known as the tangent method (see, e.g., \cite{Heimburg2007,Sugar1999, Hac2005}). To do that, a tangential line is drawn to a heat capacity curve at its low-temperature (onset of the phase transition) or high-temperature (completion of the phase transition) slopes, and the respective temperature is determined as the zero-crossing point of this tangential line. Plotting the temperatures of the onset and completion of the phase transition as a function of the composition of the lipid mixture allows one to construct an empirical experimental phase diagram of the lipid mixture.

\subsection{Monte Carlo Simulations}
\label{Sec:MCsim}
\begin{figure}[!t]
\begin{center}
\includegraphics{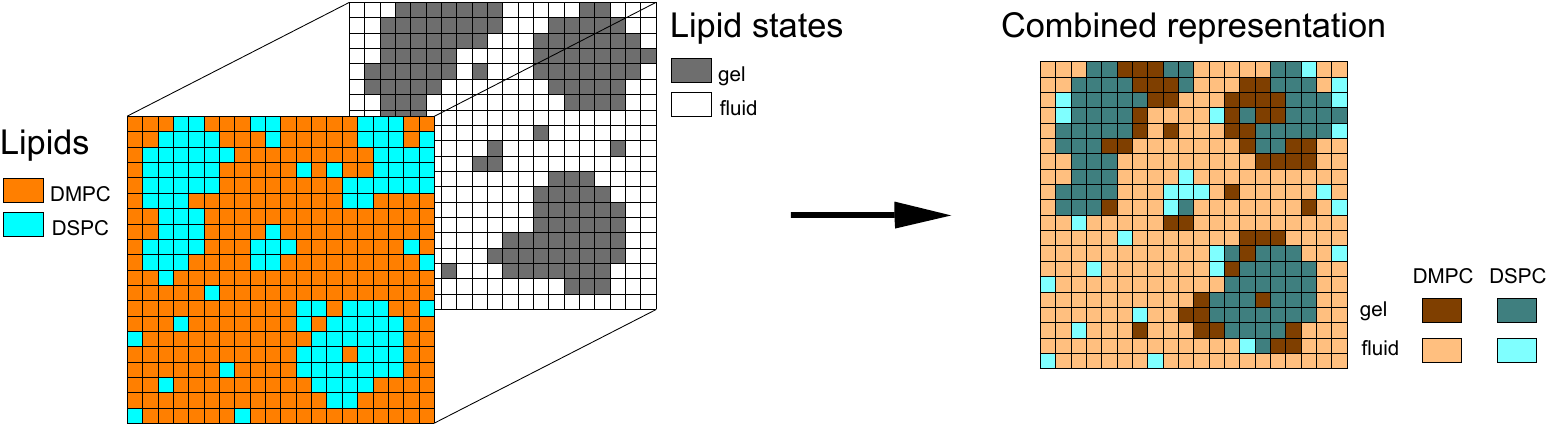}
\caption{Square lattice model used in this work to simulate lipid membranes consisting of a mixture of two lipids (DMPC and DSPC in the present study). Individual lipid molecules occupy the nods of a square lattice. Each of the two lipids (orange, cyan) can exist in either gel or fluid conformational state (black, white). The combined representation via four-colour coding makes it possible to see the complete configuration of the lattice in a single snapshot.}
\label{Fig_lattice}
\end{center}
\end{figure}

The approach to lattice-based MC simulations of a two-component membrane employed in the present study is generally similar to the one described previously \cite{Sugar1999, Hac2005, Sugar2000}. To achieve a higher computational efficiency in Monte Carlo simulations of membrane dynamics on the experimentally relevant spatial scales (of order of a micrometer) and time intervals (of order of a second), where a particular type of lipid packing and fine molecular details should be of little importance, we make one more step toward simplification of the description within the framework of the membrane lattice model. To represent a membrane, instead of a triangular lattice of lipid chains \cite{Sugar1999, Hac2005}, we use a \textit{square} lattice of lipid \textit{molecules} (figure \ref{Fig_lattice}). In this representation, neither the orientation of a lipid molecule within the membrane plane nor the states of individual lipid chains of the lipid molecule are accounted for. The model assumes that each of the lipid molecules placed on a square lattice with periodic boundary conditions can exist in either gel or fluid state, and can move laterally via next-neighbour exchange. Formally, this model is an Ising model with a conserved order parameter (representing lipids) decorated by an Ising model with a non-conserved order parameter (representing lipid states).

The fundamental step of the MC simulation consists of two sub-steps. In the first sub-step, an attempt is made to change the state of a randomly chosen lipid (from gel to fluid or vice versa). The second sub-step is an attempt to exchange the positions of a randomly chosen pair of next neighbours on the lattice.

For each sub-step the change in the Gibbs free energy
\begin{equation}
\eqalign{\fl \Delta G = \Delta N^{\rm F}_1 \left(\Delta E_1 -T\Delta S_1 \right)
							+ \Delta N^{\rm F}_2 \left(\Delta E_2 -T\Delta S_2 \right)
							+ \Delta N^{\rm GF}_{11}w^{\rm GF}_{11} \\
		 + \Delta N^{\rm GF}_{22}w^{\rm GF}_{22}
		 + \Delta N^{\rm GG}_{12}w^{\rm GG}_{12}
		 + \Delta N^{\rm FF}_{12}w^{\rm FF}_{12}
		 + \Delta N^{\rm GF}_{12}w^{\rm GF}_{12}
		 + \Delta N^{\rm GF}_{21}w^{\rm GF}_{21},}
\label{eq:Gibbs}
\end{equation}
is calculated. Here, $\Delta E_i$ and $\Delta S_i$ are the changes in the internal energy and entropy of a molecule of lipid $i$ when it switches its state from gel to fluid, $w^{mn}_{ij}$ are the next-neighbour interaction parameters of lipids $i$ and $j$ being in states $n$ and $m$, respectively, $\Delta N^{\rm F}_i$ is the change in the number of molecules of lipid $i$ in the fluid state and $\Delta N^{mn}_{ij}$ is the change in the number of next-neighbour contacts of lipids $i$ and $j$ being in states $n$ and $m$, respectively. In the simulation, the attempts of the state change and next-neighbour exchange are accepted with probability $p = 1$ for $\Delta G < 0$ and $p = \exp \left(-\Delta G/RT\right)$ for $\Delta G \geq 0$.

For an $L \times L$ lattice, one MC cycle consists of a chain of $L^2$ elementary MC steps. After each cycle, the enthalpy of the lattice is calculated as follows: $H = N^{\rm F}_1 \Delta E_1
					+ N^{\rm F}_2 \Delta E_2
					+ N^{\rm GF}_{11}w^{\rm GF}_{11}
					+ N^{\rm GF}_{22}w^{\rm GF}_{22}
					+ N^{\rm GG}_{12}w^{\rm GG}_{12}
					+ N^{\rm FF}_{12}w^{\rm FF}_{12}
					+ N^{\rm GF}_{12}w^{\rm GF}_{12}
					+ N^{\rm GF}_{21}w^{\rm GF}_{21}$.

MC simulations naturally give access to a number of thermodynamic parameters of the system. For example, the excess heat capacity $C(T)$ can be calculated from the variance of equilibrium fluctuations of the total lattice enthalpy $H$ as follows: $C(T) = {\left\langle \left(H -\langle H \rangle  \right)^2 \right\rangle}/{RT^2}$.

To adjust the lipid interaction parameters, we used the approach previously described by Sug\'{a}r \textit{et al.} \cite{Sugar1994, Sugar1999}: temperature-dependent excess heat capacity curves were obtained from MC simulations for a range of membrane compositions and compared with experimentally measured heat capacity curves, and the parameters $w^{mn}_{ij}$ were varied until a reasonable agreement with experimental DSC data was achieved. As in the previous studies \cite{Sugar1999, Hac2005}, the procedure of tuning the lipid interaction parameters results in a single set of $w^{mn}_{ij}$ which is used to describe the membrane at all temperatures and lipid compositions. Since a simpler lattice representation of the lipid system (lipid molecules arranged on a square lattice) was used in our simulations compared to the previous studies \cite{Sugar1999, Hac2005} (lipids represented as dimers of acyl chains arranged on a triangular lattice), the lipid interaction parameters $w^{mn}_{ij}$ in our study differ from those used in previous publications \cite{Sugar1999, Hac2005} (table \ref{Table1}). In the present paper, unless explicitly stated otherwise, all simulations were carried out using this set of lipid interaction parameters.

\begin{table}[!t]
\caption{Thermodynamic parameters of lattice-based MC simulations of the DMPC/DSPC lipid membranes used in the present work and previous publications \cite{Sugar1999, Hac2005}.$^{\rm a}$}
\footnotesize \rm
\begin{tabular*}{\textwidth}{@{}l@{\extracolsep\fill}c@{\extracolsep\fill}c@{\extracolsep\fill}c@{\extracolsep\fill}c@{\extracolsep\fill}c@{\extracolsep\fill}c@{\extracolsep\fill}c@{\extracolsep\fill}c@{\extracolsep\fill}c@{\extracolsep\fill}c@{}}

\br

&
\centre{10}{Thermodynamic Parameters}\\
\ns

&
\crule{10}\\

&$\Delta E_1$
&$\Delta E_2$
&$\Delta S_1$
&$\Delta S_2$
&$w^{\rm GF}_{11}$
&$w^{\rm GF}_{22}$
&$w^{\rm GF}_{12}$
&$w^{\rm GF}_{21}$
&$w^{\rm GG}_{12}$
&$w^{\rm FF}_{12}$\\

MC study
&\tiny{(${\rm J}{\rm mol}^{\textnormal{-}1}$)}
&\tiny{(${\rm J}{\rm mol}^{\textnormal{-}1}$)}
&\tiny{(${\rm J}{\rm mol}^{\textnormal{-}1}{\rm K}^{\textnormal{-}1}$)}
&\tiny{(${\rm J}{\rm mol}^{\textnormal{-}1}{\rm K}^{\textnormal{-}1}$)}
&\tiny{(${\rm J}{\rm mol}^{\textnormal{-}1}$)}
&\tiny{(${\rm J}{\rm mol}^{\textnormal{-}1}$)}
&\tiny{(${\rm J}{\rm mol}^{\textnormal{-}1}$)}
&\tiny{(${\rm J}{\rm mol}^{\textnormal{-}1}$)}
&\tiny{(${\rm J}{\rm mol}^{\textnormal{-}1}$)}
&\tiny{(${\rm J}{\rm mol}^{\textnormal{-}1}$)}\\

\mr

 \scriptsize{Ref. \cite{Sugar1999}},
&\scriptsize{12700}
&\scriptsize{21970}
&\scriptsize{42.65}
&\scriptsize{67.02}
&\scriptsize{1353}
&\scriptsize{1474}
&\scriptsize{1548}
&\scriptsize{1715}
&\scriptsize{565}
&\scriptsize{335}\\
\scriptsize{triangular lattice}&&&&&&&&&&\\
\scriptsize{of lipid chains}&&&&&&&&&&\\

 \scriptsize{Ref. \cite{Hac2005}},
&\scriptsize{13165}
&\scriptsize{25370}
&\scriptsize{44.31}
&\scriptsize{77.36}
&\scriptsize{1353}
&\scriptsize{1474}
&\scriptsize{1548}
&\scriptsize{1716}
&\scriptsize{607}
&\scriptsize{251}\\
\scriptsize{triangular lattice}&&&&&&&&&&\\
\scriptsize{of lipid chains}&&&&&&&&&&\\

 \scriptsize{Present work},
&\scriptsize{26330}
&\scriptsize{50740}
&\scriptsize{88.65}
&\scriptsize{154.70}
&\scriptsize{1827}
&\scriptsize{1622}
&\scriptsize{4025}
&\scriptsize{4460}
&\scriptsize{1412}
&\scriptsize{502}\\
\scriptsize{square lattice}&&&&&&&&&&\\
\scriptsize{of lipids}&&&&&&&&&&\\
\br

\end{tabular*}
$^{\rm a}$  Simulations by Sug\'{a}r \textit{et al.} \cite{Sugar1999} and Hac \textit{et al.} \cite{Hac2005} were carried out on the triangular lattice of lipid chains. The parameters used in the present work differ from those of \cite{Sugar1999, Hac2005} due to the fact that simulations were carried out on a square lattice of lipid molecules. Notice further, that the parameters in \cite{Sugar1999, Hac2005} are given per mole of lipid chains while in the present work they are given per mole of lipid molecules.
\label{Table1}
\end{table}

For each of the membrane compositions and temperatures addressed in the present study, simulations started with a random initial configuration of the membrane where lipids were randomly distributed on the lattice and all assigned to be in the fluid state (i.e., configurations corresponding to an infinitely high temperature). After that, the system was abruptly quenched to a particular target temperature, and simulations were carried out while keeping the temperature constant. The whole time-dependent evolution of the membrane starting with the abrupt temperature quench was recorded to study the domain coarsening and growth dynamics. To calculate the excess heat capacities and correctly address the other stationary properties of the membrane, including lipid diffusion dynamics, line tension of lipid domains, correlation lengths and susceptibilities, the membrane was first equilibrated. To achieve fast equilibration, a third substep, consisting in position exchange of two randomly chosen lipid molecules, was added to the Monte Carlo procedure. This approach, which was suggested in \cite{Sugar1997} and successfully applied in \cite{Sugar1999, Sugar2000, Michonova-Alexova2002}, is known to be very efficient in driving the system toward the equilibrium configuration. This procedure was propagated typically for $1.5 \times 10^5$ MC cycles, which for the lattice sizes used in this work is substantially longer than the typical time required by the total lattice enthalpy $H$ to reach its equilibrium value at a given temperature in the presence of the random lipid exchange substep. After the completion of this procedure, the random lipid exchange substep is switched off, and the equilibrated system is ready for studies of its equilibrium properties.

To ensure that the system has reached the full equilibrium, we verified whether the mean total lattice enthalpy relaxed to a constant value. In cases where equilibrium membrane configurations within the fluid--gel phase coexistence region of the phase diagram were required, the complete equilibration was additionally checked by visual inspection of membrane snapshots.

To represent the translational motion dynamics of lipids in the fluid and gel states in a realistic way, a rate function controlling the likelihood of next-neighbour exchange depending on the number of gel neighbours was introduced, as suggested in \cite{Hac2005}, to ensure a lower mobility of lipids in the gel environment. In particular, for two neighbouring lipids the next-neighbour exchange is accepted with a probability $r = \exp\left[-\left({N_{\rm gel~neighbours}}/{N_{\rm neighbours}}\right)\left(\Delta E/k_{\rm B}T \right) \right]$, where $N_{\rm neighbours}$ ($N_{\rm gel~neighbours}$) is the number of the next neighbours (in the gel state) for the pair of lipids attempting to exchange their positions; in our simulations on the square lattice of lipids $N_{\rm neighbours}=6$. The quantity $\Delta E \geq 0$ corresponds to the activation energy barrier for the next-neighbour exchange in the all-gel environment compared to the all-fluid environment. In our simulations, $\Delta E$ was adjusted to provide ca. 40 times slower translational diffusion of lipids in the gel phase compared to the fluid phase. Since the transition probability between the gel and fluid conformations of the lipid molecule in our simulations depends only on its immediate environment, and does not involve any time delay, it follows that at least the equilibrium properties of the membrane should not depend on the rate function.

Simulations were typically carried out on $600 \times 600$ or $400 \times 400$ lattices, which corresponds to a membrane patch consisting of $3.6 \times 10^5$ and $1.6 \times 10^5$ lipid molecules, respectively. MC simulations were run for up to $2 \times 10^7$ MC cycles, which allowed us to collect the necessary data for the analysis of thermodynamic properties of the membrane and translational diffusion of lipid molecules.
 
The spatial scale and time intervals addressed in our simulations can be obtained by relating the simulation lattice size to the lipid head group dimensions, and subsequently finding the MC step duration by relating translational diffusion of lipids to experimental data. By assuming the lipid head group having a size of ca. $0.8 \times 0.8$ nm$^2$ and using experimental data on DMPC diffusion coefficient in the fluid phase ($3 - 6$ $\mu$m$^2$/s at $T=304.5$ K \cite{Kuo1979, Vaz1985, Dolainsky1997}), we find that one MC cycle of our simulations corresponds to $\approx 50$ ns. Thus, our simulations describe the behaviour of a membrane fragment of a size of $\approx 0.48 \times 0.48$ $\mu{\rm m}^2$ over time intervals of $\approx 1$ s. These are indeed the experimentally relevant scales we planned to achieve in our simulations.

The program code is fully original and is written in Fortran completely from scratch except for the random number generation routine. The use of a good pseudo-random number generator is essential for the success of a Monte Carlo simulation study. In our simulations we therefore used the Mersenne Twister routine \cite{Matsumoto1998} providing sequences of pseudo-random numbers equidistributed in 623 dimensions characterized by an extremely long period of $2^{19937} - 1 \approx 4.3 \times 10^{6001}$. The Compaq Visual Fortran Compiler  Ver. 6.6a (Compaq Computer Corporation, Houston, TX) or the Intel Visual Fortran Compiler Professional Edition Ver. 11.1 (Intel Corporation, Santa Clara, CA) was used for compilation. The use of the Intel compiler provides a $\sim 30$\% faster performance for the same code on the same workstation. Simulation results were analyzed using original dedicated routines written in MATLAB Ver. R2007b (The MathWorks, Natick, MA) and Fortran.

Monte Carlo simulations were carried out on a workstation (Intel Core2 Quad Extreme CPU X9770 3.2 GHz, 4 GB RAM) running under Windows XP. Under these conditions, a simulation run on a $600 \times 600$ lattice for $2 \times 10^7$ MC cycles takes about 700 h (Compaq compiler) or 460 h (Intel compiler) of CPU time. The simulation method very naturally offers the opportunity of parallel implementation, which can further substantially speed-up the computation. Recent advances in the use of GPU-based computations \cite{Preis2009, Weigel2010} also offer impressive speed-up of lattice based simulations and are planned to be implemented in our future work.

\subsection{Data Analysis}

\subsubsection{Analysis of membrane configurations.}
To analyze the spatial distributions of lipids and lipid states, the radial autocorrelation function $G(r)$ and the circularly averaged structure function $S(k)$ were calculated from the radial distribution function $g(r)$ as follows \cite{Hansen2006, Fisher1964, Allen1991}
\begin{eqnarray}
G(r) = g(r) - 1, \label{eq:G}\\
g(r) = \bar{\rho}^{-2}\left\langle \sum_i \sum_{j\neq i} \delta({\bf r}_i)\delta({\bf r}_j-{\bf r})\right\rangle, \label{eq:g}\\
S(k) = 1 + \bar{\rho}\mathcal{F}\left\{G(r)\right\}, \label{eq:S}
\end{eqnarray}
where $\bar{\rho}$ is the spatially averaged particle density, and $\mathcal{F}\left\{ \cdot \right\}$ denotes the Fourier transform. The fact that the particles cannot take arbitrary positions but rather occupy sites on the square lattice is taken into account to avoid artifacts in spatial and angular averaging. When studying equilibrium membrane properties, structure functions $S(k)$ were averaged over several hundred equilibrium configurations of the membrane at a given temperature and composition.

\subsubsection{Analysis of lipid diffusion and simulation of FCS experiments.}\label{sec:MethodsFCS}
To study the lateral diffusion of lipid molecules, positions ($x$, $y$) of a small fraction ($<0.05\%$) of lipid molecules were recorded, and the time- and ensemble-averaged mean-square displacement $\textit{MSD}(\tau)$ was calculated
\begin{equation}
\textit{MSD}(\tau) = \frac{1}{t_{\rm max} - \tau} \sum^{t_{\rm max}-\tau}_{t=1}
															{ \left\{   \left[ x(t)-x(t+\tau) \right]^2
															          + \left[ y(t)-y(t+\tau) \right]^2  \right\} },
\end{equation}
where $\tau$, $t$, and $t_{\rm max}$ are integers giving time in units of MC cycles; here, $\tau$ denotes the time lag, and $t_{\rm max}$ is the total length of the lipid molecule trajectory.

Additionally, fluorescence correlation spectroscopy (FCS) \cite{Petrov2008} measurements were simulated: the tracked particles were assumed to be fluorescent, and the autocorrelation function $G(\tau) = {\left\langle \delta F(t) \delta F(t+\tau)\right\rangle}/{\left\langle F \right\rangle^2}$ of fluorescence intensity fluctuations $\delta F(t) = F(t)-\left\langle F\right\rangle$ around the mean intensity $\left\langle F\right\rangle$ in a 2D Gaussian detection volume $\exp(-2r^2/r_0^2)$ was calculated. The detection spot size $r_0$ was 31 lattice units $\approx 25$ nm, the size experimentally achievable using the stimulated emission depletion (STED) FCS technique \cite{Kastrup2005} (for more details see \cite{Ehrig2011}). FCS curves were averaged over nine different positions on the lattice and analyzed using the model
\begin{equation}
G(\tau)=G(0)\frac{1}{1+(\tau/\tau_{\rm D})^\beta},
\label{eq:FCSmodel}
\end{equation}
where the characteristic decay time $\tau_{\rm D}$ is the so-called diffusion time, which is related to the diffusion coefficient and the detection spot size (in case of normal diffusion, $\tau_{\rm D} = r_0^2/(4D)$), and the exponent $\beta$ is used to describe  the deviation of the autocorrelation curve from the normal diffusion model. For $\beta=1$ expression (\ref{eq:FCSmodel}) corresponds to normal diffusion, while for $0 < \beta < 1$ it provides a simple way to describe anomalous subdiffusion in FCS \cite{Petrov2008}.

\section{Results}

\subsection{Phase Diagram of the DMPC/DSPC system based on heat capacity data}

\begin{figure}[!t]
\begin{center}
\includegraphics{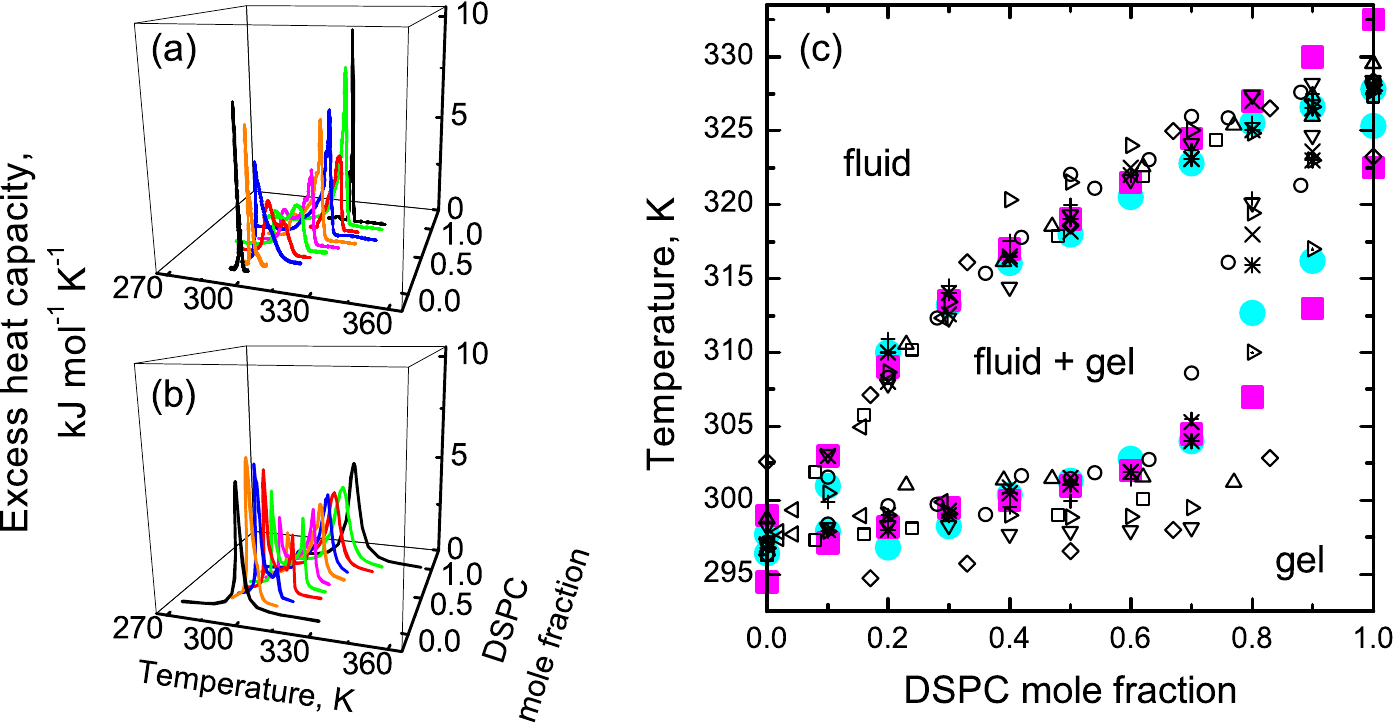}
\caption{(a, b) Excess heat capacity curves for DMPC/DSPC lipid membranes for a range of compositions (DMPC/DSPC = 0/100, 10/90, 20/80, ..., 90/10, 100/0) measured experimentally by differential scanning calorimetry (a) and obtained in our MC simulations (b). (c) Empirical phase diagram of DMPC/DSPC constructed from the excess heat capacity curves. Experimental data: present work (\fullcircle), \cite{Shimshick1973} (\opensquare), \cite{Mabrey1976} (\opencircle), \cite{Lentz1976} (\opendiamond), \cite{Dijck1977} (\opentriangledown), \cite{Wilkinson1979} ($\triangleleft$), \cite{Nibu1995} (\opentriangle), \cite{Sugar1999, Sugar2000} ($\triangleright$), our reanalysis of experimental $C(T)$ data for DMPC/DSPC 20/80 and 10/90 from \cite{Sugar2000} ($\triangleright$\hspace{-1.6mm}$\cdot$\hspace{0.5mm}), and \cite{Hac2005} ($\times$). Monte Carlo simulation data: present work (\fullsquare), \cite{Sugar1999} ($+$), \cite{Hac2005} ($\ast$).}
\label{Fig_CT_PhaseDiagram}
\end{center}
\end{figure}

The experimentally measured heat capacity curves in figure \ref{Fig_CT_PhaseDiagram}(a) capture the well-known feature of two-component lipid mixtures (see, e.g., \cite{Heimburg2007}), namely, that in a two-lipid system with a fixed composition the all-fluid and all-gel states of the membrane are separated by a temperature range where coexistence of the fluid and gel phases takes place. This temperature range can span several tens of degrees. By contrast, phase coexistence is not observed in single-lipid systems, which, upon cooling down below the phase transition temperature, directly undergo a transition from the fluid to gel state.

After successfully adjusting the interaction parameters $w^{mn}_{ij}$ for the simulation of DMPC/DSPC on the square lattice, we find that our MC simulations reproduce well the experimentally measured excess heat capacity curves $C(T)$, as shown in figure \ref{Fig_CT_PhaseDiagram}(a, b). 

Historically, in the lipid studies, one usually speaks about broadening of the transition in a two-lipid system (here, by transition, one means the fluid--gel transition, and the broadening is understood as compared to a single-lipid system), and the onset and completion temperatures determined from experimental data are usually discussed, with a fluid--gel phase coexistence region located between these temperatures \cite{Sugar1999, Shimshick1973, Wilkinson1979}. Going beyond the empirical experimental approach, one finds that the transition from a single-phase state of the membrane to a two-phase coexistence state is actually a phase transition by itself. A two-lipid system upon cooling from a high temperature at which it is in the fluid state, would thus undergo two phase transitions: from the fluid state to the fluid--gel coexistence, and, upon further cooling, another phase transition from the fluid--gel coexistence to the gel state. To avoid confusion, we clarify that, wherever the expression ``phase transition'' in the present paper is applied to a two-lipid system, the phase transition from a single phase state into a two-phase coexistence state (or vice versa) is referred to. Studying the properties of these phase transitions, we believe, can reveal many details on the microscopic organization of the lipid membrane.

Heat capacity curves $C(T)$ (both obtained experimentally and in MC simulations) were analyzed using the tangent method (see, e.g. \cite{Heimburg2007}) to determine the onset and completion temperatures (i.e., estimates of the temperature of the transition from a single-phase state to two-phase coexistence). To do that, the outer slopes of the $C(T)$ profiles for a range of compositions (DMPC/DSPC = 0/100, 10/90, 20/80, ..., 90/10, 100/0) were fit with straight lines passing through the corresponding inflection points. The onset and completion temperatures in this case are defined as the respective intersections of the tangential lines with the zero-line. As a result, an experimental phase diagram for the binary lipid mixture can be constructed. We point out that in our work we apply the tangent method to analyze $C(T)$ data for all membrane compositions, inculding the single-lipid systems. This method results in a finite transition width for pure DMPC and DSPC membranes. The transition widths of single-lipid systems obtained in our MC simulations are in agreement with previous lattice-based MC simulation results \cite{Sugar1994}, and as previously the simulated $C(T)$ peaks of individual lipids are wider than the ones observed experimentally -- for discussion see \cite{Sugar1994}. Notice that, in contrast to the present work, the onset and completion temperatures for the single lipids are frequently not reported in the literature, but rather a single temperature corresponding to the peak of the $C(T)$ dependence is given in the phase diagram.

As is evident from figure \ref{Fig_CT_PhaseDiagram}(c), our MC simulation-based data are in perfect agreement with experimental results (both obtained in the present work and published earlier by other groups \cite{Sugar1999, Hac2005, Shimshick1973, Mabrey1976, Lentz1976, Dijck1977, Wilkinson1979, Nibu1995}), as well as with results from the previous simulation studies  \cite{Sugar1999, Hac2005} carried out on the triangular lattice of lipid chains. This shows that the particular choice of a lattice geometry for lattice-based MC simulations should not be important, provided the lipid interaction parameters are chosen in a proper way.\footnote[1]{While the arguments in favour of representing a lipid membrane as a triangular lattice of lipid chains may seem more or less sound in case of the gel state of the membrane, they do not look nearly as convincing in case of the fluid state. Moreover, there exists experimental evidence that the arrangement of lipid chains in the gel phase shows only a short-range positional order and thus represents the hexatic phase \cite{Kjaer1987, Smith1990, Bernchou2009}, and therefore the triangular lattice of chains does not correctly represent the structure of the gel phase. This seriously weakens the arguments in favour of the triangular lattice of chains as the only correct choice in lattice-based lipid membrane simulations.}

It should be pointed out that generally there is neither proof nor guarantee that the onset and completion temperatures extracted from the $C(T)$ curves by the tangent method necessarily correspond to the actual temperatures for the phase transitions from the gel state to fluid--gel coexistence and from fluid--gel coexistence to the fluid state, respectively. Therefore, the \textit{empirical} phase diagram constructed on the basis of heat capacity data may differ from the real phase diagram of the system and thus not provide an insight into the microscopic structure and the dynamics of the membrane.

\subsection{Phase separation behaviour of the DMPC/DSPC system. Binodals and spinodals}
\begin{figure}[!t]
\begin{center}
\includegraphics{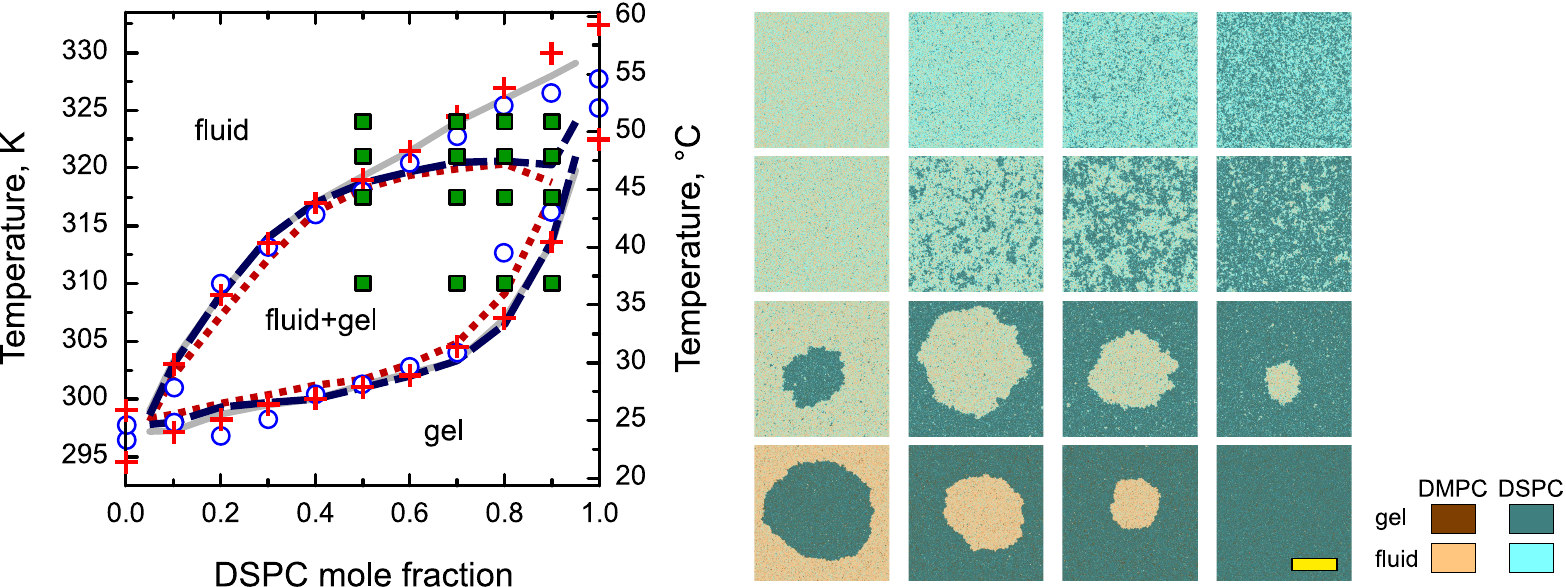}
\caption{Left-hand panel: Phase diagram of the DMPC/DSPC lipid mixture. Empirical results obtained from the analysis of excess heat capacity curves from differential scanning calorimetry measurements (\opencircle) and MC simulations ($+$). Lipid state spinodal (\dashed), lipid state binodal (\longbroken) and lipid demixing curves (\full) are shown. Right-hand panel: Membrane configurations corresponding to the compositions and temperatures marked by filled squares on the phase diagram. Lattice size: $600 \times 600$; scale bar: 200 lattice units $\approx 160$ nm.}
\label{Fig_PhaseDiagram_snapshots}
\end{center}
\end{figure}

A closer look at the membrane snapshots from the MC simulation at different compositions and temperatures, as shown in figure \ref{Fig_PhaseDiagram_snapshots}, indeed reveals that the phase behaviour is, in fact, more complicated. In the region bounded by the curves of the onset and completion temperatures, as determined from the experimental and simulated  $C(T)$ data (circles and crosses in figure \ref{Fig_PhaseDiagram_snapshots}), two types of phase behaviour can be observed: i) complete separation of the gel and fluid phases and ii) a highly dynamic intermixing of the two phases. To gain better understanding of this phenomenon, the radial autocorrelation function $G(r)$ (\ref{eq:G}) and the structure function $S(k)$ (\ref{eq:S}) were calculated for a wide range of membrane compositions and broad temperature intervals covering the whole area of the phase diagram. Importantly, our simulations allow us to study $G(r)$ and $S(k)$ for the spatial distributions of both lipids (DMPC or DSPC) and their states (fluid or gel) independently, thereby providing us with valuable details on the microscopic lipid organization in the membrane and its role in the dynamics of the phase separation.

We found that outside the region with a clear coexistence of two phases (the exact boundaries of this region will be discussed in what follows), structure functions of lipid states $S_{\rm S}(k)$ are well described by the Ornstein--Zernike (OZ) approximation \cite{Fisher1964} $S_{\rm S}(k) = S^{\rm OZ}(k) = {S_0}/{\left[ 1 + (\xi k)^2\right]}$, where $k$ is a wavenumber, $\xi$ is the correlation length of fluctuations, and the amplitude $S_0$ is proportional to the order parameter susceptibility of the system.\footnote[2]{The order parameter susceptibility characterizes the sensitivity of the order parameter in the system to the external perturbation. For example, for a liquid--gas transition the order parameter is the local density in the system, and the role of the order parameter susceptibility is played by the isothermal compressibility; in case of the Ising model consisting of magnetic spins, one deals, correspondingly, with the magnetic susceptibility. In the membrane system in question, the local order parameter corresponds to the lipid conformation, and the order parameter susceptibility will, among other things, characterize the efficiency of lipid domain formation around proteins preferentially wetted by one of the membrane phases (see \cite{Ehrig2011} and refs. therein).}

The analysis of the structure functions in terms of the OZ approximation allowed us to study the temperature dependences of the correlation length $\xi_{\rm S}(T)$ and the amplitude $S_{{\rm S},0}(T)$ of the structure function, and thus to determine the lipid state spinodal and binodal.

The binodal curves were determined from the condition $d(1/\xi_{\rm S}(T))/dT = 0$ \cite{Strobl1997}. The estimated accuracy of the binodal determination was found to be $\pm 1$ K. Notice that, when extended to DMPC/DSPC 0/100 and 100/0 compositions, the upper and lower binodal curves are expected to meet at the respective lipid melting transition temperatures, which is in agreement with the data we obtained (see figure \ref{Fig_PhaseDiagram_snapshots}). The origin of the nonmonotonic behavior of the upper binodal at DSPC mole fractions exceeding 85\% still needs to be explained.

The spinodals can be determined by extrapolating the $1/S_{{\rm S},0}(T)$ dependence to zero crossing \cite{Strobl1997}. To determine the position of the upper (lower) spinodal from the results of our simulations, a set of $S_{{\rm S},0}(T)$ data above the upper binodal (or below the lower binodal, respectively) was fit for every membrane composition to a dependence $1/S_{{\rm S},0}(T) = B |1-T/T_{\rm sp}|^p$, where $B$ is a constant prefactor, $T_{\rm sp}$ is the spinodal temperature, and $p = \mathcal{O}(1)$. To do that, typically $7-14$ data points spanning a temperature range of $5-10$ K were used. Note that for a system at criticality, the exponent $p$ is nothing but the critical exponent $\gamma$ (see section \ref{sec:Critical}). The  accuracy in determination of spinodals ranges from $\pm 1$ K in the regions where they approach close to binodals to $\pm 3$ K in the regions where the spinodals depart from the binodal curves.

Surprisingly, we found that the phase diagram based on the binodal and spinodal curves differs significantly from the empirical phase diagram obtained from the heat capacity data (figure \ref{Fig_PhaseDiagram_snapshots}).

\begin{figure}[!t]
\begin{center}
\includegraphics{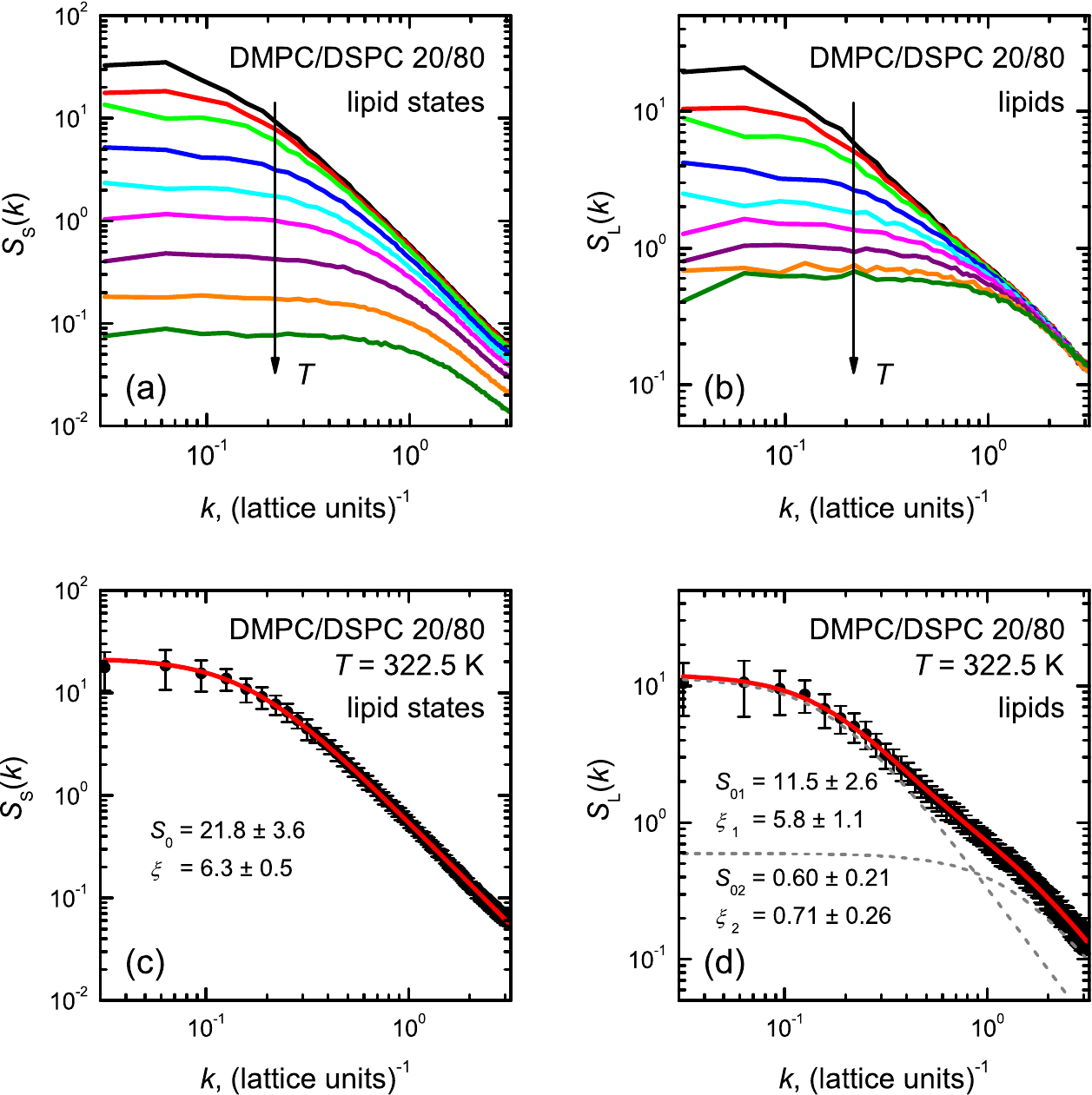}
\caption{Representative structure functions characterizing the spatial distribution of lipid states $S_{\rm S}(k)$ (a) and lipids $S_{\rm L}(k)$ (b) for a set of temperatures above the phase transition (top to bottom: $T=$ 322, 322.5, 323, 324, 325, 326, 328, 331, and 335~K). Weighted least-squares fit of $S_{\rm S}(k)$ at $T=$ 322.5~K using the single-component OZ approximation (c); weighted least-squares fit of $S_{\rm L}(k)$ at $T=$ 322.5~K using the two-component OZ approximation (d). The grey dashed lines in (d) show the single components of the fit. For clarity, $S(k)$ data are shown without the constant offset.}
\label{Fig_StructureFactors}
\end{center}
\end{figure}

What is the reason for such a strong difference? To answer this question, we analyzed the structure functions $S_{\rm L}(k)$  characterizing the spatial distribution of lipids in the membrane outside the phase coexistence region. Quite unexpectedly, it turned out that the structure functions of lipids cannot be adequately described by the simple OZ approximation, and two OZ components are necessary to provide an adequate fit of the structure functions of lipids: $S_{\rm L}(k) = S_{\rm L1}^{\rm OZ}(k)+S_{\rm L2}^{\rm OZ}(k)$ (figure \ref{Fig_StructureFactors}).

It should be noted that in order to provide a perfect fit of the $S_{\rm S}(k)$ and $S_{\rm L}(k)$ data within the whole range of wavenumbers, it was required that a small positive offset had to be included into the corresponding fit functions. This offset, being important only at high $k$-numbers, is clearly a consequence of carrying out simulations on a finite-size lattice, and was consistently reproduced in our simulations of the Ising model on the square lattices of the corresponding sizes (data not shown).

\begin{figure}[!t]
\begin{center}
\includegraphics{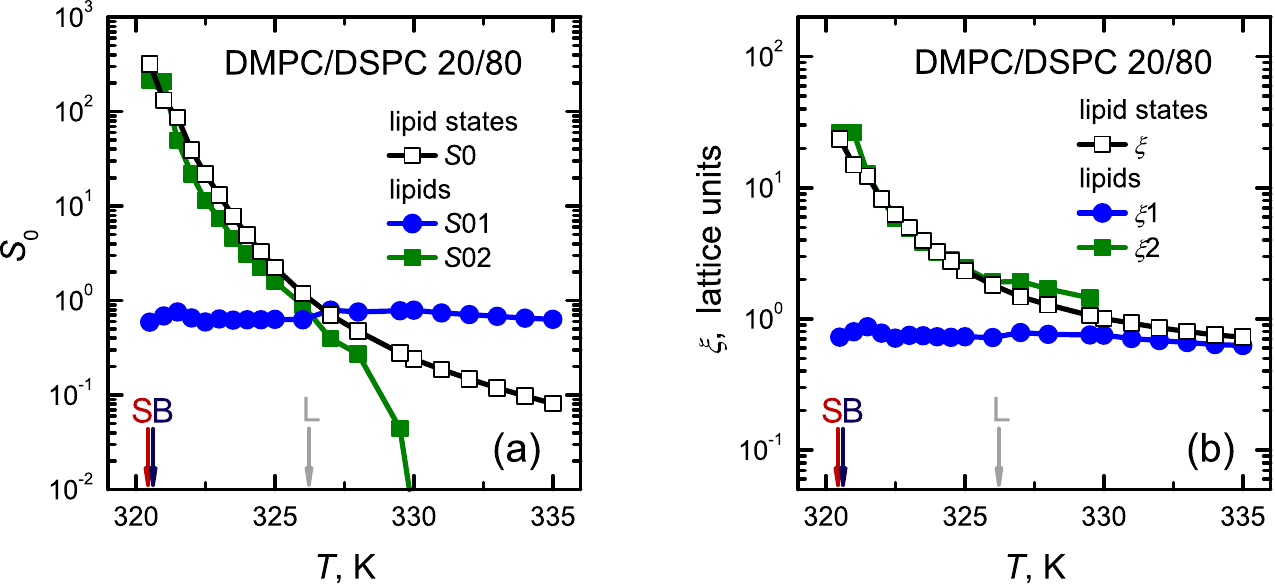}
\caption{Temperature dependences of the amplitudes (a) and correlation lengths (b) of the OZ components for lipids and lipid states of the DMPC/DSPC 20/80 mixture. The arrows indicate the positions of the binodal (B), spinodal (S), and lipid demixing (L) temperatures.}
\label{Fig_Sk_analysis}
\end{center}
\end{figure}

It appears that parameters $S_{\rm L1}^{\rm OZ}(0)$ and $\xi_{\rm L1}$ of the first OZ-component only weakly depend on temperature and correspond to demixing of lipids in the same state; $S_{\rm L2}^{\rm OZ}(0)$ and $\xi_{\rm L2}$, on the other hand, show strong temperature dependences similar to those of $S_{\rm S}(0)$ and $\xi_{\rm S}$ (figure \ref{Fig_Sk_analysis}(a, b)) and thus reflect the appearance of dynamic domains.

We found that for each lipid composition, the temperature at which the amplitudes of the two OZ components used to describe the structure function of the spatial distribution of lipids are equal, $S_{\rm L1}^{\rm OZ}(0) = S_{\rm L2}^{\rm OZ}(0)$, perfectly matches the temperature corresponding to the peak in the $C(T)$ profile. Physically, this means that at this temperature the susceptibilities responsible for the spontaneous clustering of molecules of the same lipid (irrespectively whether they are in the fluid or gel conformation) and their clustering into a particular membrane phase, become equal, and the peak in the heat capacity, in fact, marks the intense local lipid demixing due to transition from a single phase to a two-phase coexistence state. Therefore, we termed the curve defined by the condition $S_{\rm L1}^{\rm OZ}(0) = S_{\rm L2}^{\rm OZ}(0)$ as the \textit{lipid demixing} curve. Notice that the lipid demixing curve very closely approximates the empirical phase diagram based on the heat capacity data (figure \ref{Fig_PhaseDiagram_snapshots}). Therefore, in what follows we will refer to the boundaries of the empirical phase diagram as heat capacity-based lipid demixing curves.

We point out that in the region of the phase diagram corresponding to the gap between the binodal and the lipid demixing curve, the membrane shows behaviour characteristic of neither the single-phase nor two-phase coexistence states. Namely, on the one hand, the membrane is not uniform anymore as in the fluid state, while on the other hand, the fluid and gel domains are very poorly defined, have sizes on a wide range of spatial scales, and show non-stop fluctuations without any signs of nucleation and growth.

In fact, this behaviour is characteristic of near-critical fluctuations expected to take place in the vicinity of a critical point. This suggestion is further supported by the behaviour of the binodal and spinodal curves which, within the limits of our accuracy, touch in the corresponding region of the phase diagram, which should take place at the critical point. We will come again to the issue of the critical behaviour of the membrane in more detail later, in sections \ref{sec:Scenarios} and \ref{sec:Critical}.

Here we remark that in an earlier MC simulation-based study of a DMPC/DSPC membrane \cite{Michonova-Alexova2002}, a conclusion was made based on the analysis of size distributions of fluid and gel clusters, that the phase diagram of this lipid system is much more complicated than that based on heat capacity data (the work \cite{Michonova-Alexova2002}, however, did not discuss the possibility of the presence of a critical point in the system). Although the quantitative results of the above paper may suffer from a very small system size addressed ($40 \times 40$ lipid chains), on the qualitative side, these conclusions are largely in agreement with the ones we draw in the present study using different arguments.

\subsection{Transient lipid domains versus thermodynamically equilibrium phases}
\label{sec:DomainsVsPhases}

It has been pointed out elsewhere \cite{Heimburg2007} that, when discussing phase separation in lipid membranes, one should make a clear distinction between the thermodynamically stable phases and (usually, small) transient membrane domains statistically appearing and disappearing in an equilibrated membrane.

In this context, we would like to check to what extent the concept of the thermodynamically equilibrium phase can be applied to results of our simulations. This is important in view of the apparent controversy regarding the applicability of the lever rule to the phase separation in two-component lipid membranes: on the one hand, the recent laser scanning microscopy experiments on two-component membranes (GUVs) \cite{Fidorra2009b} show that the gel and fluid domains in two-component membranes do indeed correspond to equilibrium phases, and therefore the lever rule applies to their description, on the other hand, in previously published MC simulation studies \cite{Sugar1999, Sugar2000} on the DMPC/DSPC systems the authors found out that the fluid and gel phase in the two-phase coexistence region do not follow the lever rule (this was interpreted in \cite{Sugar1999, Sugar2000} as being a consequence of the fact that the gel--fluid transition in DMPC/DSPC mixtures is not a first-order phase transition).

To resolve this contradiction, we studied whether the binodal curves which were determined in the previous section of the paper, can be successfully reconstructed by determining for different fixed membrane compositions the temperature dependence of the composition of the individual \textit{phases}, which should exactly be the case if the lever rule applies to the system.

Once one can clearly distinguish between the fluid and gel phases, the problem becomes trivial and is reduced to determining the relative amounts of the two lipids in each of the assigned membrane phases. In practice, of course, this should be carried out using only equilibrium membrane configurations, and averaging over a number of configurations is required to obtain reliable results.

\begin{figure}[!t]
\begin{center}
\includegraphics{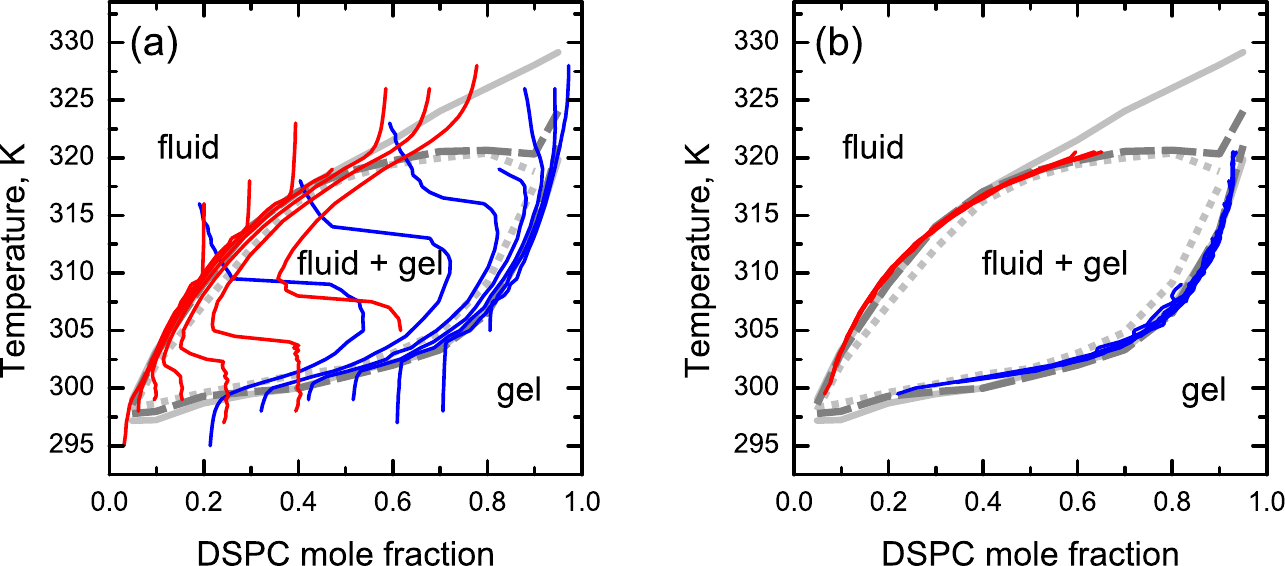}
\caption{DMPC/DSPC phase diagram reconstructed from the analysis of the temperature-dependent lipid compositions of the fluid and gel phases. Red and blue curves show the compositions (DSPC mole fraction) of the fluid and gel phases, respectively, as a function of temperature for different membrane compositions. (a) Direct calculation from equilibrium membrane configurations. (b) Calculation after selecting the gel and fluid phases by segmentation of the lipid state configurations (see text for details). Additionally, the lipid state binodal (long-dashed curves) and spinodal (short-dashed curves), and lipid demixing curves (grey solid curves) are shown (cf. figure \ref{Fig_PhaseDiagram_snapshots}).}
\label{Fig_lever}
\end{center}
\end{figure}

In our analysis, we first assumed, as it has been done in \cite{Sugar1999}, that the fluid and gel phase in a particular membrane configuration are identified with the conformational state of the lipid: all lipids in the fluid conformation are counted as belonging to the fluid phase, and, correspondingly, all lipids in the gel conformation are counted as belonging to the gel phase. The results obtained in this way are presented in figure \ref{Fig_lever}(a). Clearly, this analysis fails to consistently reproduce the binodals of the system, which is in qualitative agreement with previous results of \cite{Sugar1999}.

Our interpretation of the reason why this analysis fails is different from the one previously proposed in \cite{Sugar1999}. A careful examination of membrane configurations allowed us to conclude that this analysis fails due to the presence of small short-lived fluid or gel domains, which, because of their transient nature, cannot be considered as a part of the thermodynamically stable phase. If counted as belonging to a particular phase, these domains can severely distort the phase ratio, as well as the lipid composition of the lipid and gel phases, and lead to a break-down of the lever rule.

Therefore, one can hope that, if the small transient membrane domains are excluded from the analysis, and only the thermodynamically stable membrane phases are taken into account, the same analysis will reproduce the binodal curves of the system with a better accuracy.

The problem, thus, boils down to selecting an appropriate rule to discriminate between small transient membrane domains and thermodynamically stable phases. We found that this could be successfully accomplished by low-pass spatial filtering of the lipid state configurations of the membrane. In particular, we found that a procedure consisting of applying a 2D Gaussian filter with the standard deviation of $\sigma = 5$ lattice units and subsequent thresholding the result at the $50\%$ level resulted in consistent reproducible results.\footnote[3]{An approach based on time averaging of membrane configurations over reasonably short simulation intervals was found to give similar results.} This procedure applied to results of MC simulations with lipid compositions DMPC/DSPC in the range from 20/80 to 80/20, consistently reproduced the upper and lower binodals of the system (figure \ref{Fig_lever}(b)).

The above procedure used for detection of fluid or gel phases in the membrane eliminates from the analysis small domains which contain below $\sim 50$ lipid molecules, which is in agreement with the intuitive understanding that a portion of a membrane that can be classified into a thermodynamically stable phase should consist of a number of molecules much larger than one. Our present analysis shows that this number is of order of $10^2$. Domains of this size ($\lesssim 10 \times 10$ nm$^2$) are clearly below the optical resolution and are not observable in confocal fluorescence microscopy experiments, which additionally have a finite exposure time, thereby eliminating features fluctuating on a short time scale.

Thus, the apparent controversy between the recent fluorescence microscopy data \cite{Fidorra2009b} and previously published results of MC simulations \cite{Sugar1999} is resolved if one takes into consideration only relatively large and stable membrane domains which constitute the thermodynamically stable phases, which is certainly the case for the domains detectable in fluorescence microscopy experiments.

\subsection{Different scenarios of phase transition}
\label{sec:Scenarios}

A careful examination of the membrane simulation snapshots at various temperatures and lipid compositions shows that, depending on the membrane composition, the phase transition from the fluid phase to the fluid--gel phase coexistence can take place according to two difference scenarios.

\begin{figure}[!p]
\begin{center}
\includegraphics{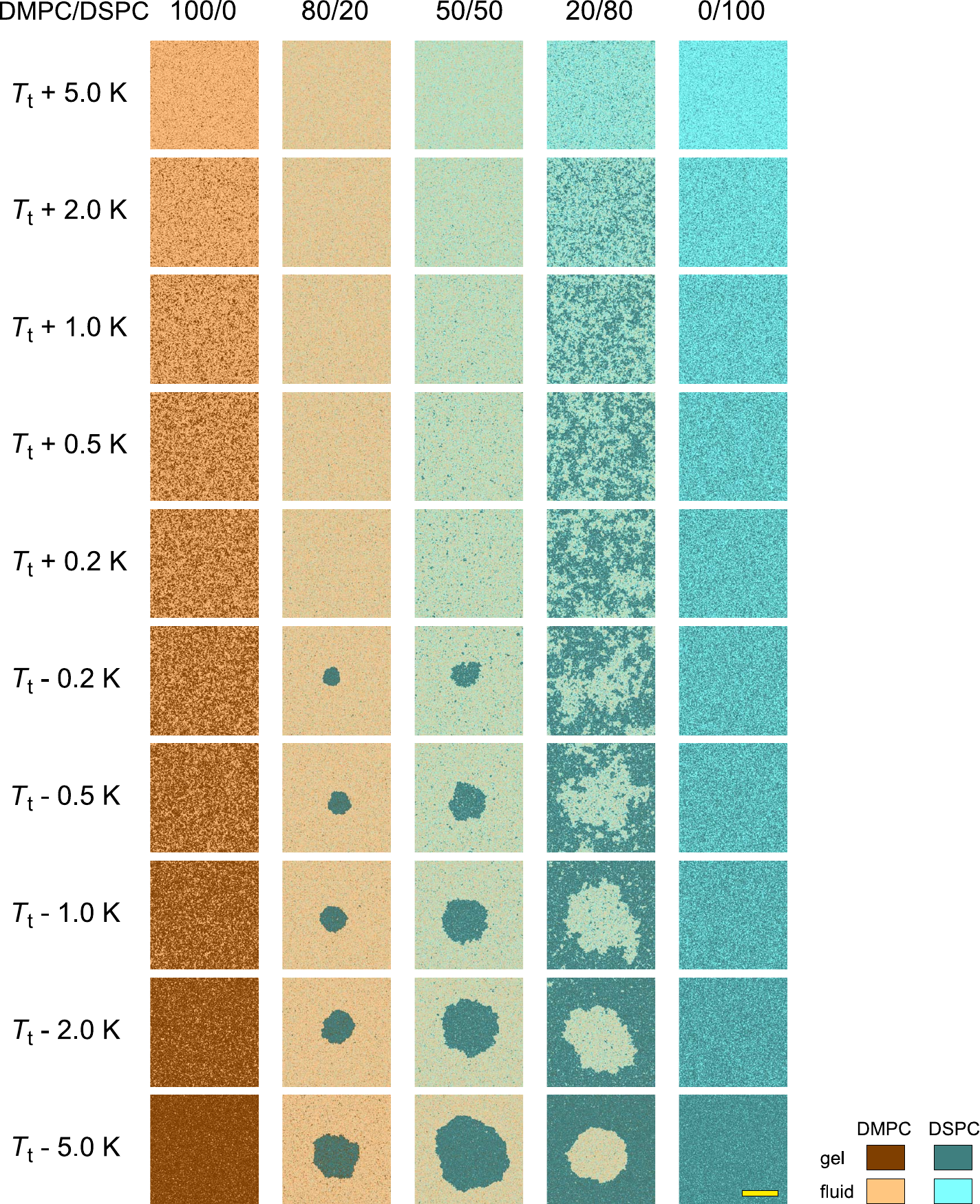}
\caption{Representative equilibrium configurations obtained in MC simulations of the DMPC/DSPC system with three different compositions DMPC/DSPC = 80/20, 50/50, and 20/80 for a set of temperatures relative to the phase transition from the fluid state to fluid--gel coexistence. For comparison, results for the pure DMPC and DSPC membranes depicting their fluid--gel transitions are shown. Transition temperatures for DMPC/DSPC 100/0, 80/20, 50/50, 20/80 and 0/100 are $T_{\rm t}$ = 297.0, 309.7, 318.7, 320.5, and 328.0 K, respectively. Lattice size: $600 \times 600$; scale bar: 200 lattice units $\approx 160$ nm. See text for a discussion of the different  phase transition scenarios.}
\label{Fig_PhaseTransSnapshots}
\end{center}
\end{figure}

At compositions for which the lipid state binodal coincides with the lipid demixing curve (see figure \ref{Fig_PhaseDiagram_snapshots}), the transition has a quasi-abrupt character. This is exemplified in figure \ref{Fig_PhaseTransSnapshots} by the corresponding temperature sequences of equilibrated membrane snapshots for DMPC/DSPC 80/20 and 50/50. There, lowering the temperature below the transition temperature $T_{\rm t}$ immediately leads to an abrupt fluid--gel phase separation and formation of a circular-shaped, stable domain. Above $T_{\rm t}$, the membrane stays in the all-fluid phase.

On the other hand, in the region of the phase diagram where the lipid demixing curve strongly deviates from the lipid state binodal, a completely different scenario is observed. In particular, when the membrane is cooled down to temperatures below the lipid demixing curve, transient random-shaped fluctuating domains appear. As can be seen from the corresponding snapshots for DMPC/DSPC 20/80 in figure \ref{Fig_PhaseTransSnapshots}, with lowering the temperature, these domains become larger, but remain fractal-shaped and highly dynamic. The mean size of these gel and fluid domains, as well as the order-parameter susceptibility, which are determined from the analysis of the structure function $S_{\rm S}(k)$, diverge when approaching the temperature of phase transition (see section \ref{sec:Critical}). This means that, compared to the scenario discussed above, the character of the transition is changed, and we are dealing with a continuous phase transition. We find that within the corresponding range of lipid compositions, the lipid state spinodal closely approaches the binodal. Eventually these curves touch at a common (local) maximum, i.e. at the point where the compositions of the fluid and gel phase become indistinguishable, which should take place at a critical point. The shapes of the binodal and spinodal curves suggest that the critical composition of the membrane should be close to DMPC/DSPC 20/80. To determine the position of the critical point more exactly, we construct a curve in the phase diagram along which the amounts of lipids in the fluid and gel states are equal: $X_{\rm fluid} = X_{\rm gel} = 0.5$. At the critical point one should generally expect that not only the compositions of the fluid and gel phase become indistinguishable, but also their relative contributions become equal.

\begin{figure}[!t]
\begin{center}
\includegraphics{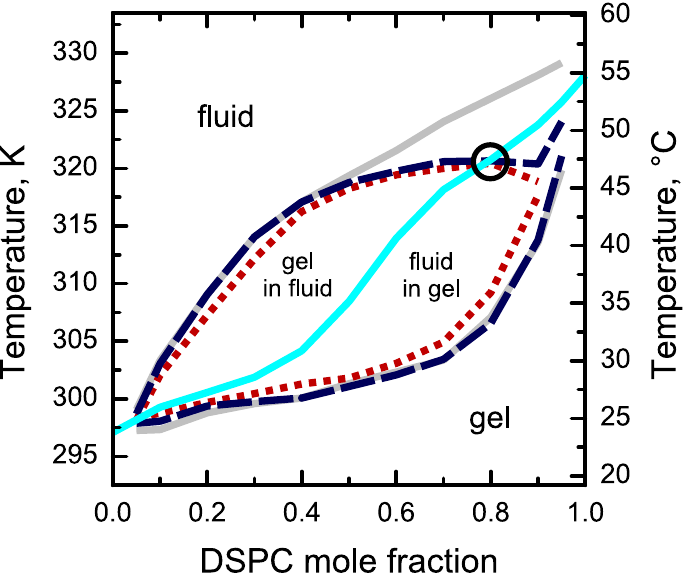}
\caption{Phase diagram of the DMPC/DSPC system demonstrating the presence of a critical point. Lipid state spinodal (short-dashed lines), lipid state binodal (long-dashed lines) and lipid demixing curves (solid grey lines) are the same as in figure \ref{Fig_PhaseDiagram_snapshots}. The solid cyan line marks temperatures at which the DMPC/DSPC membrane shows equal amounts of the fluid and gel phase $X_{\rm fluid} = X_{\rm gel} = 0.5$. The critical point is marked with an open circle. Within the two-phase fluid--gel coexistence region, areas marked as ``gel in fluid'' and ``fluid in gel'' correspond to the gel and fluid minority phases, respectively. Upon complete equilibration, the minority phase always forms a single circular-shaped domain surrounded by the majority phase.}
\label{Fig_CriticalPoint}
\end{center}
\end{figure}

Indeed, as can be clearly seen from figure \ref{Fig_CriticalPoint}, this curve crosses the binodal and spinodal lines exactly at the point where they touch, which confirms the presence of a critical point at this position. Thus, in the region of the phase diagram where the lipid demixing curve strongly deviates from the lipid state binodal, approaching the phase transition is accompanied by near-critical fluctuations of the fluid and gel phases. Cooling the membrane to temperatures below the phase transition leads to large-scale phase separation, and a single large domain of a fluid or gel phase is eventually formed in a fully equilibrated membrane. Notice that in the vicinity of the critical point, these domains show strong thermally induced shape fluctuations due to a decrease of the line tension in the vicinity of the critical point (see discussion in section \ref{Sec:LineTension}).

The existence of critical points in lipid membranes has been suggested in several recent experimental studies. In particular, (near-)critical behaviour was observed in giant unilamellar vesicles of ternary lipid mixtures \cite{Honerkamp-Smith2008} and in giant plasma membrane vesicles (with a lipid composition close to the one of the cell plasma membrane) \cite{Veatch2008}. Remarkably, the behaviour observed in \cite{Veatch2008} is very similar to the one in our MC simulation: depending on the membrane composition, the transition to the two-phase coexistence takes place either in an abrupt manner -- when the membrane does not pass through a critical point, or via critical fluctuations -- when the membrane does pass through a critical point (compare our figure \ref{Fig_PhaseTransSnapshots} with figure 1(d, e) of \cite{Veatch2008}).

Furthermore, there is experimental evidence of the existence of a critical point in the binary DMPC/DSPC lipid mixture. Atomic force microscopy experiments \cite{Nielsen2007} provided structural evidence of criticality in DMPC/DSPC monolayers. More interesting conclusions were made \cite{Knoll1981, Knoll1983, Sackmann1995b} for DMPC/DSPC bilayers, where the existence of a critical point hidden within the fluid--gel coexistence region of the phase diagram has been suggested based on small-angle neutron scattering experiments. Notice that the critical point we find in our MC simulation is also located within the region of the phase diagram, where, based on the calorimetry data, the fluid--gel coexistence would be generally expected.


In addition, we arrive at quite an interesting conclusion: namely, that the binodals of the system are best reproduced via the analysis of the lipid compositions of the fluid and gel phases (see section \ref{sec:DomainsVsPhases}) exactly at those membrane compositions and temperatures where the phase transition in the lipid system takes place in a quasi-abrupt manner, i.e., when it is close by its properties to the first-order transition, that is exactly under conditions where this analysis is expected to hold.

In case of a single-lipid membrane, the phase transition also takes place via spatial fluctuations of the fluid and gel domains (see corresponding sequences of membrane snapshots in figure \ref{Fig_PhaseTransSnapshots}). For a single-lipid system, according to the Gibbs' phase rule, no phase coexistence can take place either above or below the transition temperature. In agreement with that, the results of MC simulations for single lipids in the vicinity of the phase transition show transient fluid or gel domains, similar to what was observed in a recent atomic-scale molecular dynamics simulation study \cite{Murtola2006}. Similar to \cite{Murtola2006} we find that at $T = T_{\rm t} + 8$~K these small domains typically exist for no longer than $\sim 100$ ns. Even at the transition temperature the domains are transient and their sizes are relatively small ($\sim 100$ lipids for DMPC and $\sim 20$ lipids for DSPC) and, importantly, do not change with an increase of the simulation lattice size. Remarkably, this is in a very good agreement with results of a previous MC simulation study of a DPPC membrane using a ten-state Pink model \cite{Cruzeiro-Hansson1988}, where the mean cluster size at the transition temperature was estimated to be $\sim 40$ lipids, in agreement with experimental estimates \cite{Freire1978}. Interestingly, this also very well agrees with our estimate of the size of transient lipid clusters we obtained above in section \ref{sec:DomainsVsPhases}.

Before concluding this section, three important points should be emphasized.

First, equilibration of the membrane within the fluid--gel phase coexistence region in our simulations always ultimately results in formation of a \textit{single} domain of the minority phase, surrounded by the continuous majority phase. In fact, this should be expected in this system, because in the phase coexistence region of the phase diagram the system evolution is driven so as to minimize the line tension energy, and, since no explicit or implicit penalties on the domain growth are imposed in the present simulation, a single minority phase domain should eventually be formed. On the other hand, in cases where domains of the minority phase show a different curvature compared to the majority phase, the curvature mismatch can result in such a penalty, which at some point stops the growth of domains and prevents their coalescence (see, e.g., \cite{Baumgart2003, Auth2009a, Ursell2009, semrau2009, Idema2010}). Therefore, the presence of a number of minority phase domains in a two-phase coexistence region in a lattice-based simulation which does not penalize the domain growth and/or coalescence should generally mean that this membrane configuration is still far from the equilibrium one, and therefore, any results on diffusion of lipids (by either studying the time dependence of the mean-square displacement or simulating results of fluorescence correlation spectroscopy (FCS) experiments) obtained in this regime have nothing to do with the equilibrium properties of the membrane.

Second, we remark on the shape of the domains of the minority phase formed as a result of phase separation after the membrane is abruptly quenched from $T=\infty$ to a temperature within the fluid--gel coexistence. For the minority phase content $X < 1/\pi$ the domains always have a circular shape. For larger fractions of the minority phase, a single circular domain is also initially formed. Notice, however, that if simulations are carried out, as in the present paper, on a 2D square area with periodic boundary conditions (toroidal boundary conditions), and the fraction of the minority phase is in the range of $1/\pi < X < 1/2$, the complete equilibration of the system should eventually lead to the formation of a stripe-shaped domain, whereas a circular domain represents an extremely long-lived metastable state \cite{Neuhaus2003, Fischer2010}. For simulations on the $L \times L$ lattice, the time $\tau$ required for the conversion of a circular domain to a stripe-shaped domain in case of the Ising system strongly depends on the system size \cite{Neuhaus2003}: $\tau \propto L^2 \exp(2 s \lambda L/k_{\rm B}T)$, where $s = 0.1346...$ and $\lambda$ is the phase interface line tension (for determination of line tension in our system, see section \ref{Sec:LineTension}). Our preliminary numerical experiments showed that the behaviour of our system is consistent with the above expression, in spite of the fact that our system is more complicated than the Ising model. It should be emphasized that for large enough lattices the time $\tau$ becomes extremely long: we estimate it as $\tau \sim 10^{10}$~MC cycles for $L = 400$ and $\tau \sim 10^{13}$~MC cycles for $L = 600$.

Moreover, we believe that the stripe shapes of minority phase domains, which are expected at the minority phase content $1/\pi < X < 1/2$ in lattice-based simulations with periodic (toroidal) boundary conditions, are, in fact, of little importance when simulations are directly compared with experimental data: according to recent results based on off-lattice simulations \cite{Fischer2010}, the formation of a band-shaped domain is just an artifact of the toroidal boundary conditions, and is not observed on a surface of a sphere if the particle interactions are isotropic. The latter observation generally agrees with experiments on phase separation in giant unilamellar vesicles. In a similar way, circular domains are also expected for supported lipid bilayers (which obviously do not have periodic boundary conditions).

Note that the present discussion is restricted exclusively to the effects of boundary conditions in simulation studies and is \textit{not} related to experimental findings for two-component membranes, where, depending on the membrane tension, either circular-shaped or stripe-shaped domains can be observed \cite{Li2006}. There, the appearance of stripe-shaped domains is most likely related to the anisotropic orientational distribution of lipid molecules induced by the high membrane tension.

Third, outside the fluid--gel phase coexistence region, small transient domains can spontaneously form and afterwards spontaneously dissolve after very short time intervals. These domains have a typical diameter of $\sim 10$ lipids or smaller, and, importantly, their average size does not change with an increase in the size of the simulation lattice.  Because of their extremely small size and short lifetime, their presence should be of no importance for optical microscopy-based experiments, as well as for simulations of these experiments at the experimentally relevant spatial scales and time intervals.

\subsection{(In)Sensitivity of the phase diagram to perturbation of lipid interaction parameters}
\label{sec:parameters}
In this section we will demonstrate that the main features of the phase diagram obtained in our MC simulations are stable with respect to a specific choice of the lipid interaction parameters. In particular, the presence of the critical point in the phase diagram is not affected by reasonable perturbations of the interaction parameters.

A straightforward approach to this issue would involve a systematic study of the phase diagram in full detail for a number of differently perturbed sets of the interaction parameters, i.e. determining binodals, spinodals and lipid demixing curves, as well as checking the presence of near-critical fluctuations and determining the position of the critical point. This approach, however, would involve immense computational efforts.

As it was discussed in the previous section, away from the critical point, the phase transition takes place in an abrupt manner, whereas in the vicinity of the critical point, the approach to the phase transition takes place via near-critical fluctuations (see figure \ref{Fig_PhaseTransSnapshots}). Therefore, to study the stability of phase diagram with respect to perturbation of the set of interaction parameters on the qualitative level it should be enough to check whether the phase transition takes place abruptly or via the near-critical fluctuations,  for respective membrane compositions, independent of reasonable perturbations of the lipid interaction parameters. One should generally expect that if the presence of the critical point is stable with respect to reasonable perturbations of the  $w^{mn}_{ij}$, the membrane composition and temperature characterizing the critical point should not change much, and a behaviour similar to the one depicted in figure \ref{Fig_PhaseTransSnapshots} for each of the lipid compositions should also be observed with the perturbed interaction parameters.

As discussed in section \ref{Sec:MCsim}, the adequate description of the experimental heat capacity data by the MC model requires to determine the lipid interactions parameters $w^{mn}_{ij}$ with the accuracy of a few percent. Therefore, it seems reasonable to carry out the stability analysis by introducing 10\% perturbations to the parameters $w^{mn}_{ij}$.

To simplify the notation, we combine the parameters $w^{mn}_{ij}$ in a single vector ${\bf W} = \{w^{\rm GF}_{11}, w^{\rm GF}_{22}, w^{\rm GG}_{12}, w^{\rm FF}_{12}, w^{\rm GF}_{12}, w^{\rm GF}_{21}\}$. Then results of simulations for the unperturbed parameter set $\bf{W}$ should be compared with the results obtained for several perturbed parameter sets ${\bf W} + {\boldsymbol \Delta}^{(k)}$. The components of the perturbation vector ${\boldsymbol \Delta}^{(k)}$ were defined as follows: $\Delta^{(k)}_i = \delta \zeta^{(k)}_i W_i$, where ${\boldsymbol \zeta}^{(k)}$ is a (random) vector whose components take values $\pm 1$, and $\delta = 0.1$ is the relative perturbation level.

Simulations were carried out for three lipid compositions DMPC/DPSC 80/20, 50/50, and 20/80 for a set of temperatures $T_{\rm t} + \Delta T$, $\Delta T = -3, -2, -1, -0.5, +0.5, +1, +2, +3$~K for the respective compositions, using four different perturbation vectors ${\boldsymbol \Delta}^{(k)}$, $k = 1, 2, 3, 4$, characterized by the following ${\boldsymbol \zeta}^{(k)}$: ${\boldsymbol \zeta}^{(1)} = \{ +1, -1, +1, -1, -1, +1 \}$, ${\boldsymbol \zeta}^{(2)} = \{ +1, -1, -1, -1, +1, +1 \}$, ${\boldsymbol \zeta}^{(3)} = \{ -1, +1, -1, +1, +1, -1 \}$, and ${\boldsymbol \zeta}^{(4)} = \{ -1, +1, +1, +1, -1, -1 \}$.

\begin{figure}[!t]
\begin{center}
\includegraphics{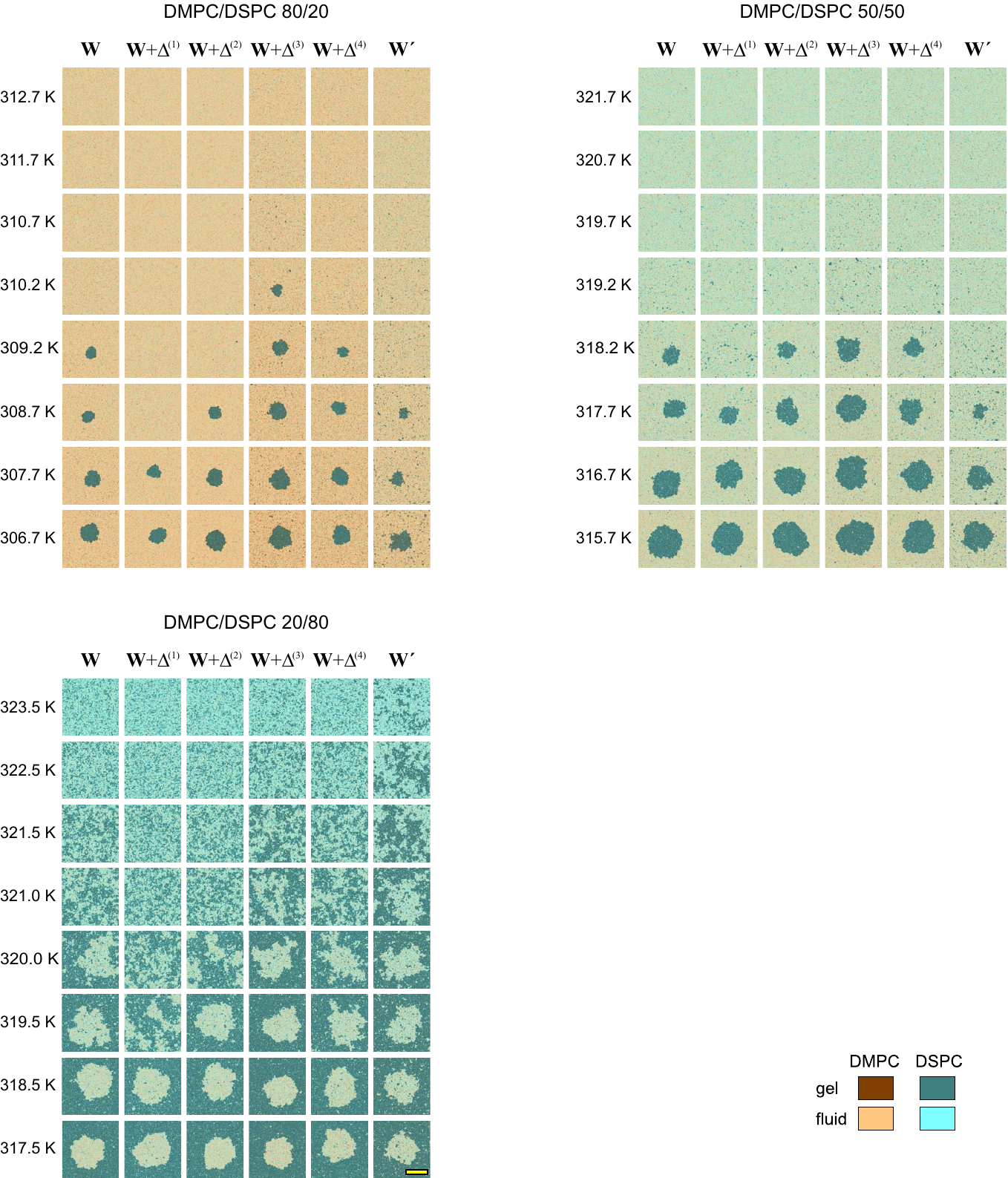}
\caption{Equilibrium membrane configurations for three lipid compositions DMPC/DSPC 80/20, 50/50, and 20/80 obtained with the unperturbed set of lipid interaction parameters ${\bf W}$, four different perturbed sets of lipid interaction parameters ${\bf W} + {\boldsymbol \Delta}^{(k)}$, $k = 1, 2, 3, 4$, and the alternative set of parameters ${\bf W'}$ (see text). The results are presented for the set of temperatures $T_{\rm t} + \Delta T$, $\Delta T = -3, -2, -1, -0.5, +0.5, +1, +2, +3$~K for the respective compositions. Transition temperatures for DMPC/DSPC 80/20, 50/50, and 20/80 are $T_{\rm t}$ = 309.7, 318.7, and 320.5 K, respectively. See text for details. Lattice size: $400 \times 400$; scale bar: 150 lattice units $\approx 120$ nm.}
\label{Fig_ParameterDependence}
\end{center}
\end{figure}

As it is clear from figure \ref{Fig_ParameterDependence}, although the 10\% perturbations in the interaction parameters can shift the phase transition temperature up or down by up to 2 K and slightly change the relative amount of the fluid and gel phases, the character of the transitions at these three lipid compositions remains essentially the same. Namely, while for DMPC/DSPC 80/20 and 50/50 the transition remains abrupt, the near-critical fluctuations accompanying the phase transition for DMPC/DSPC 20/80, irrespectively of perturbation of the parameters, indicate that the phase diagram and its critical point are stable with respect to the reasonable perturbations of the lipid interaction parameters.

We find that varying the single-lipid interaction parameters (within reasonable limits) strongly affects the width of the $C(T)$ peak for single lipids; in particular, choosing these parameters lower than the optimum will lead to broadening of the single-lipid $C(T)$ peaks. On the other hand, even with a non-optimal choice on non-optimal single-lipid interaction parameters, the rest of the parameter set can be chosen so that $C(T)$ profiles for mixtures of DMPC/DSPC within the range 10/90 to 90/10 are well reproduced, which allows one to reproduce well the main features of the empirical phase diagram.

A thorough reanalysis of our results, that we carried out recently after the main body of our study was finished, allowed us to conclude that our original choice of singe-lipid interaction parameters was sub-optimal, which lead to broadening of single-lipid $C(T)$ peaks in our MC simulations. Nevertheless, as mentioned above, this set of parameters very well reproduces the phase diagram for membrane compositions ranging from 10/90 to 90/10. Adjusting the single-lipid interaction parameters to reproduce the experimentally observed widths of $C(T)$ peaks of single lipids along with readjustment of the two-lipid interaction parameters has resulted in the following set of lipid interaction parameters ${\bf W'} = \{w'^{\rm GF}_{11}, w'^{\rm GF}_{22}, w'^{\rm GG}_{12}, w'^{\rm FF}_{12}, w'^{\rm GF}_{12}, w'^{\rm GF}_{21}\} = \{2030, 2211, 1356, 435, 3174, 3002\}{\rm ~J~mol}^{-1}$.

We have found that with this choice of interaction parameters provides additionally excellent description of the empirical phase diagram also outside the range 10/90 to 90/10 (data not shown). On the other hand, we found that the characteristic behaviour of the system was not changed if the parameter set ${\bf W'}$ was used instead of ${\bf W}$. Similar to what is observed with the parameter set ${\bf W}$, the use of the parameter set ${\bf W'}$ leads to the near-critical fluctuations in the upper right part of the phase diagram and quasi-abrupt phase transitions elsewhere. This is demonstrated in figure \ref{Fig_ParameterDependence}, where the rightmost columns show the temperature behaviour of the membrane simulated using the parameter set ${\bf W'}$. A preliminary analysis shows that changing from the lipid interaction parameter set ${\bf W}$ to ${\bf W'}$, while not affecting the qualitative picture, does lead to certain quantitative changes: in particular, we believe that, if the parameter set ${\bf W'}$ is used, the critical point is shifted closer to the lipid composition DMPC/DSPC 15/85 with the critical temperature close to 322.5~K.

\subsection{Anomalous diffusion of lipid molecules due to near-critical fluctuations}

In our recent work \cite{Ehrig2011}, using the same MC simulation and lipid system as in the present study, we showed that near-critical fluctuations can lead to transient anomalous subdiffusion of lipid molecules in a membrane. We concluded there that this phenomenon should occur irrespectively of particular details of the system except for the proximity to the critical point. Here, we demonstrate that, indeed, also for the alternative set of parameters ${\bf W'}$ discussed in section \ref{sec:parameters}, the transient anomalous subdiffusion can be observed when the system is in a state of near-critical fluctuations.

\begin{figure}[!t]
\begin{center}
\includegraphics{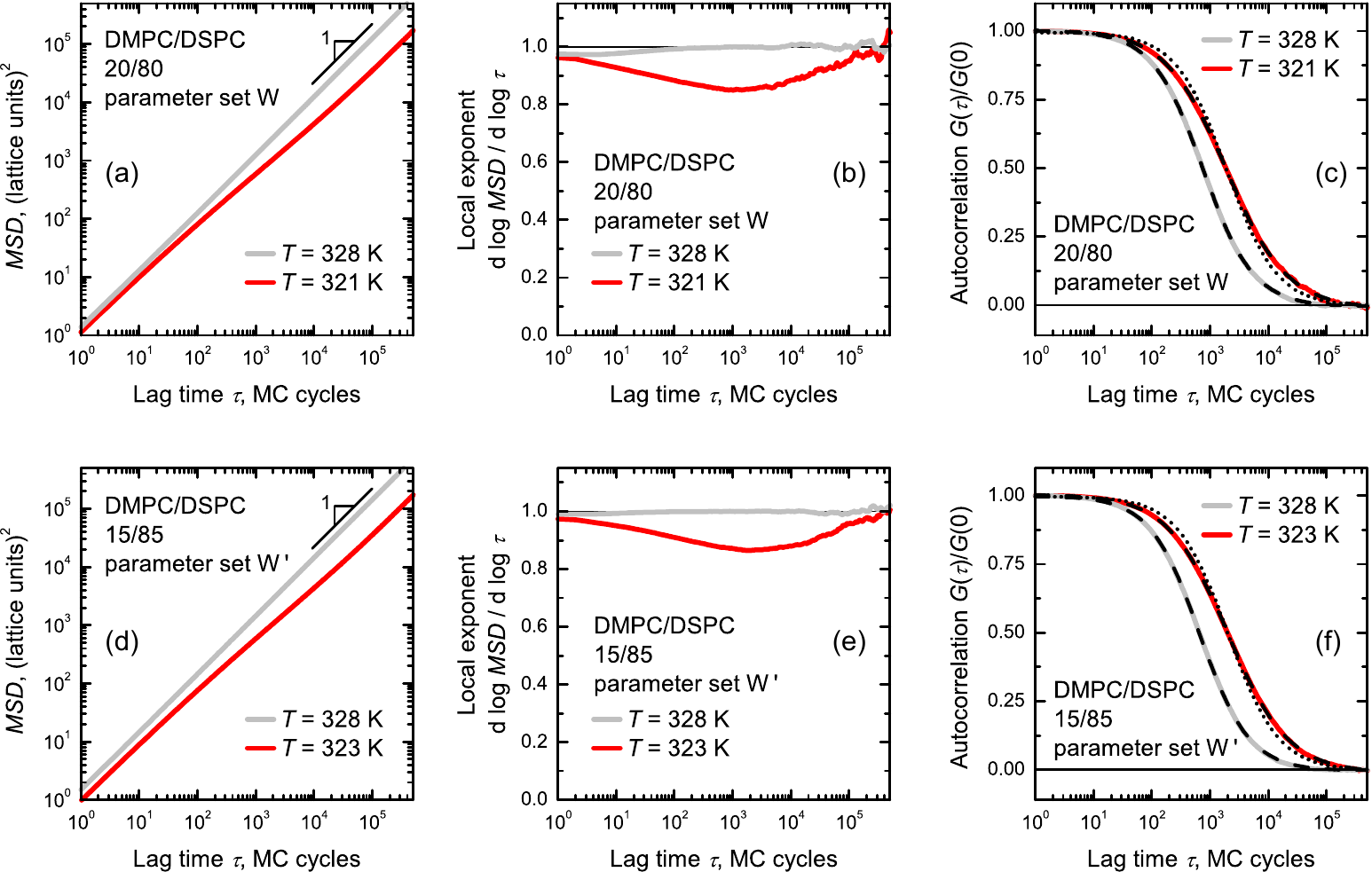}
\caption{Transient anomalous diffusion of DMPC lipids in a DMPC/DSPC membrane close to the critical point for two different sets of lipid interaction parameters. Mean square displacement $\textit{MSD}(\tau)$ (a, d), local exponent $d\log {\textit{MSD}(\tau)} /d \log \tau$ (b, e), and FCS autocorrelation $G(\tau)/G(0)$ (c, f) for the two sets ${\bf W}$ (a--c) and ${\bf W'}$ (d--f) of lipid interaction parameters (see text). Red curves show results of simulations carried out at DMPC/DSPC 20/80, $T = 321$~K (a-c) and DMPC/DSPC 15/85, $T = 323$~K (d-f) -- the temperatures and compositions close to the critical point for the respective parameter sets. Grey curves show results of simulations carried out at the same respective membrane compositions but at $T = 328$~K, away from the critical point. FCS results (c, f) are shown along with fits using the FCS diffusion model (\ref{eq:FCSmodel}) (\longbroken) -- for DMPC/DSPC 20/80 (c) this yields $\beta = 1.01$ for  $T = 328$~K and $\beta = 0.86$ for $T = 321$~K; for DMPC/DSPC 15/85 (f) $\beta = 1.02$ for  $T = 328$~K and $\beta = 0.84$ for $T = 323$~K. For comparison, fits with the fixed $\beta = 1.0$ are shown for DMPC/DSPC 20/80, $T = 321$~K (c) and DMPC/DSPC 15/85, $T = 323$~K (f)(\dotted).}
\label{Fig_Diffusion}
\end{center}
\end{figure}

As already discussed above, the near-critical behaviour in the membrane is preserved when changing the parameter set from $\bf{W}$ to $\bf{W'}$. It appears that the critical point is slightly shifted and is located close to DMPC/DSPC 15/85, $T = 322.5$~K. We studied whether the diffusion of DMPC molecules in a DMPC/DSPC 15/85 mixture at $T=323$~K (very close to the critical point) and $T=328$~K (away from the critical point) shows the same behaviour as the one described in \cite{Ehrig2011}. The MSD curves in figure \ref{Fig_Diffusion}(a, d) along with the plots of their local exponents $d\log {\textit{MSD}(\tau)} /d \log \tau$ (figure \ref{Fig_Diffusion}(b, e)) clearly show that the qualitative behavior is very similar for the two parameter sets. Namely, that near-critical fluctuations in the membrane can lead to transient anomalous subdiffusion that extends over several orders of magnitude in time while away from the critical point, diffusion is normal. This is also reflected by the FCS autocorrelation curves presented in figure  \ref{Fig_Diffusion}(c, f). In agreement with the MSD data, at high temperatures, away from criticality, the diffusion is normal. When the system is approaches criticality the appearance of transient anomalous diffusion behaviour dominates the presented FCS curves, which can only be successfully described using the model for anomalous diffusion. In particular, fits using (\ref{eq:FCSmodel}) result in $\beta = 0.86$ (figure \ref{Fig_Diffusion}(c), DMPC/DSPC 20/80, $T=321$~K, parameter set $\bf{W}$) and $\beta = 0.84$ (figure \ref{Fig_Diffusion}(f), DMPC/DSPC 15/85, $T=323$~K, parameter set $\bf{W'}$).

\subsection{Line tension}
\label{Sec:LineTension}
The importance of line tension for phase separation in lipid membranes has been shown experimentally for both liquid--liquid coexistence in ternary lipid mixtures \cite{Honerkamp-Smith2008, Baumgart2003, semrau2009, Esposito2007, Camley2010, Baumgart2005, Tian2007, Garcia-Saez2007} and for fluid--gel coexistence in binary lipid mixtures \cite{Li2006, deAlmeida2002, Leidy2002, Blanchette2007, Connell2006, Gordon2006}. The importance of line tension in fluid--gel phase separation was pointed out already 15 years ago in a lattice-based simulation study \cite{Joergensen1995}, although the study did not address the line tension in a quantitative way. Recently, line tension between the gel and fluid phases has been addressed in non-lattice-based simulations \cite{Marrink2005, Shi2005}.

As discussed above, in our simulations the DMPC/DSPC mixtures at compositions and temperatures corresponding to the fluid--gel phase coexistence region of the phase diagram always display circular-shaped domains, which can be attributed to the presence of an effective line tension between the fluid and the gel phase. After equilibration the system shows complete phase separation, and only a single domain of the minority phase remains surrounded by the majority phase, which corresponds to minimization of the perimeter of the phase interface, which is driven by minimization of the line tension energy.

As already discussed in section \ref{sec:parameters}, for membranes with the fraction of the minority phase in the range of $1/\pi < X < 1/2$ the true equilibrium state is characterized by a stripe-shaped domain while the circular domain represents a metastable state. On the other hand, for the large lattice sizes we addressed in this study, this metastable state is extremely long-lived and thus the circular-shaped domains persist for the whole simulation run.

\begin{figure}[!t]
\begin{center}
\includegraphics{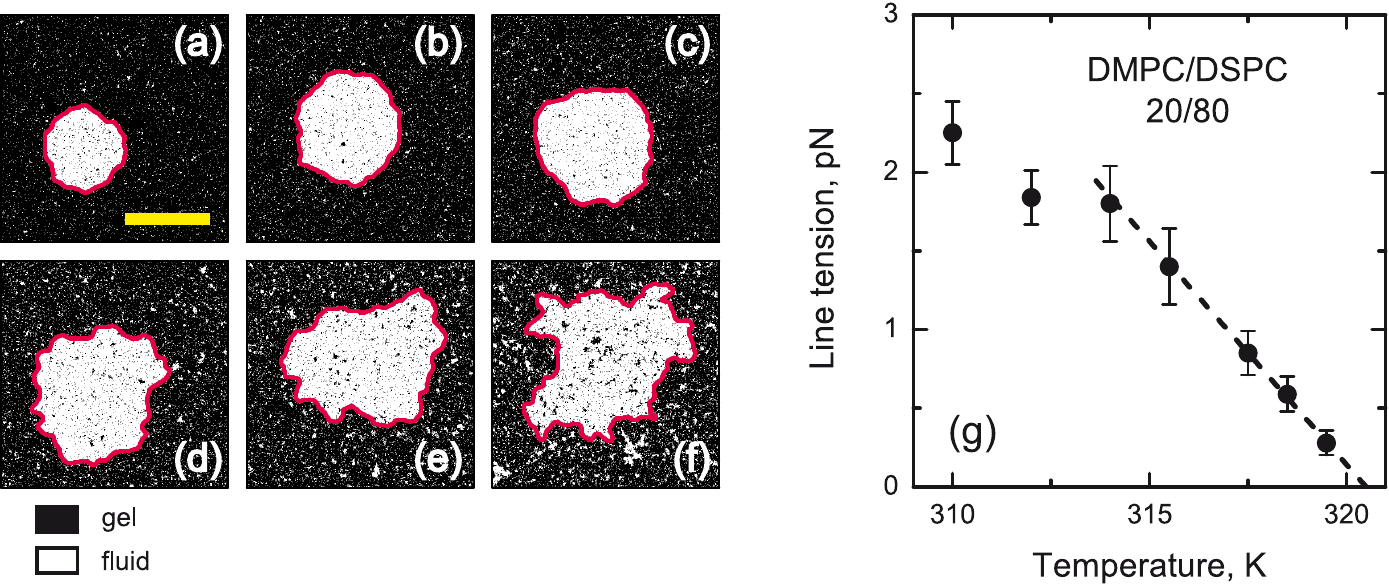}
\caption{(a-f) Representative equilibrium membrane configurations of DMPC/DSPC 20/80 membrane at $T = 310$ (a), 314 (b), 315.5 (c), 317.5 (d), 318.5 (e), and 319.5 K (f) obtained in MC simulations. The fluid and gel phases are shown as white and black, respectively. The determined domain boundaries are shown in red (see text for details). Lattice size: $400 \times 400$; scale bar: 150 lattice units $\approx 120$ nm. (g) Line tensions calculated from the Fourier spectrum of domain contour fluctuations (\fullcircle) and a linear fit of the high-temperature part of these data (\dashed) giving an estimate of the critical temperature $T_{\rm c} = (320.5 \pm 0.4)$ K.}
\label{Fig_LineTension}
\end{center}
\end{figure}

These membrane domains show thermally-excited shape fluctuations, as it can be clearly seen in the corresponding membrane snapshots in figures \ref{Fig_PhaseDiagram_snapshots}, \ref{Fig_CriticalPoint}, and \ref{Fig_LineTension}(a--f). An analysis of fluctuating domain shapes should in principle allow one to determine the line tension between the fluid and gel phases in the membrane by analyzing a sequence of fluctuating domain contours. The line tension is then determined from the Fourier spectrum of domain contour fluctuations, using the approach originally introduced by Goldstein and Jackson \cite{Goldstein1994} and later successfully applied to lipid bilayer membranes \cite{Baumgart2003, Esposito2007} (for corrected expressions, see \cite{Camley2010}). One of the important requirements of such an analysis is conservation of the domain area. Unlike the lipid composition of the membrane, which is kept strictly constant during an MC simulation, the relative amount of the gel and fluid phase at a fixed temperature and membrane composition in an equilibrated membrane is conserved only on average. As a result, the domain area is also subject to fluctuations. We have found, however, that the magnitude of these fluctuations is rather low (typically, below $3 \%$), which allows one to consider the domain area as effectively constant and thus to extract the line tension from the analysis of fluctuating domain contours.

To determine the fluctuating contours of a membrane domain, the snapshots of the membrane (reflecting only the conformational state of lipids) were first segmented by applying low-pass filtering  (convolution with a 2D Gaussian with $\sigma = 5$ lattice units) followed by thresholding of the result. A set of contours (typically, $200-500$ contours) was analyzed using a procedure described in \cite{Esposito2007, Camley2010} to extract the energies of the Fourier modes of domain fluctuations, and ultimately the values of the line tension corresponding to these modes. Since the low-pass filtering procedure employed for domain contour determination suppresses the contributions of the higher-order fluctuation modes, only the first $m$ modes satisfying the condition $m < {\pi \left\langle R \right\rangle}/{(6\sigma)}$, where $\left\langle R \right\rangle$   is the mean domain radius, were used for determination of the line tension. This typically amounted to the line tension analysis based on the first 10 to 20 modes, depending on the domain size.

We found that away from the critical point the line tension of membrane domains obtained from our simulation data is about 2 pN (figure \ref{Fig_LineTension}(g)). This value is in agreement with experimental values of a few pN reported for membranes with various lipid compositions, including ternary lipid systems exhibiting fluid--fluid phase coexistence \cite{Honerkamp-Smith2008, Baumgart2003, semrau2009, Esposito2007, Camley2010, Baumgart2005, Tian2007, Garcia-Saez2007} and binary lipid mixtures showing fluid--gel coexistence \cite{Blanchette2007}. Moreover, the line tension of 2 pN is in good agreement with the value expected \cite{Garcia-Saez2007} for the experimentally measured height mismatch of $\sim 0.9-1.3$ nm between the gel and fluid phases in DMPC/DSPC membranes \cite{Giocondi2004, Giocondi2001}; it also agrees well with the value obtained in another binary lipid system showing fluid--gel coexistence with a similar hight mismatch \cite{Blanchette2007}.

When the membrane temperature and composition approach the critical point, the line tension approaches zero (figure \ref{Fig_LineTension}), as it is generally expected \cite{Widom1965} (see more details below). On the other hand, for lipid compositions displaying a quasi-abrupt phase transition (i.e., compositions with the DSPC fraction of $\sim 50 \%$ and below in the high-temperature part of the phase diagram) the line tension does not gradually approach zero as the phase transition temperature is approached (data not shown), which is in agreement with the behaviour illustrated in figure \ref{Fig_PhaseTransSnapshots}.

\subsection{Behaviour of the membrane in the vicinity of the critical point}
\label{sec:Critical}

When approaching the critical point, many parameters of a thermodynamic system will either tend to zero or diverge as some power of $|T - T_{\rm c}|$ \cite{Fisher1964, Hohenberg1977}. These powers are known as critical exponents.

\begin{figure}[!t]
\begin{center}
\includegraphics{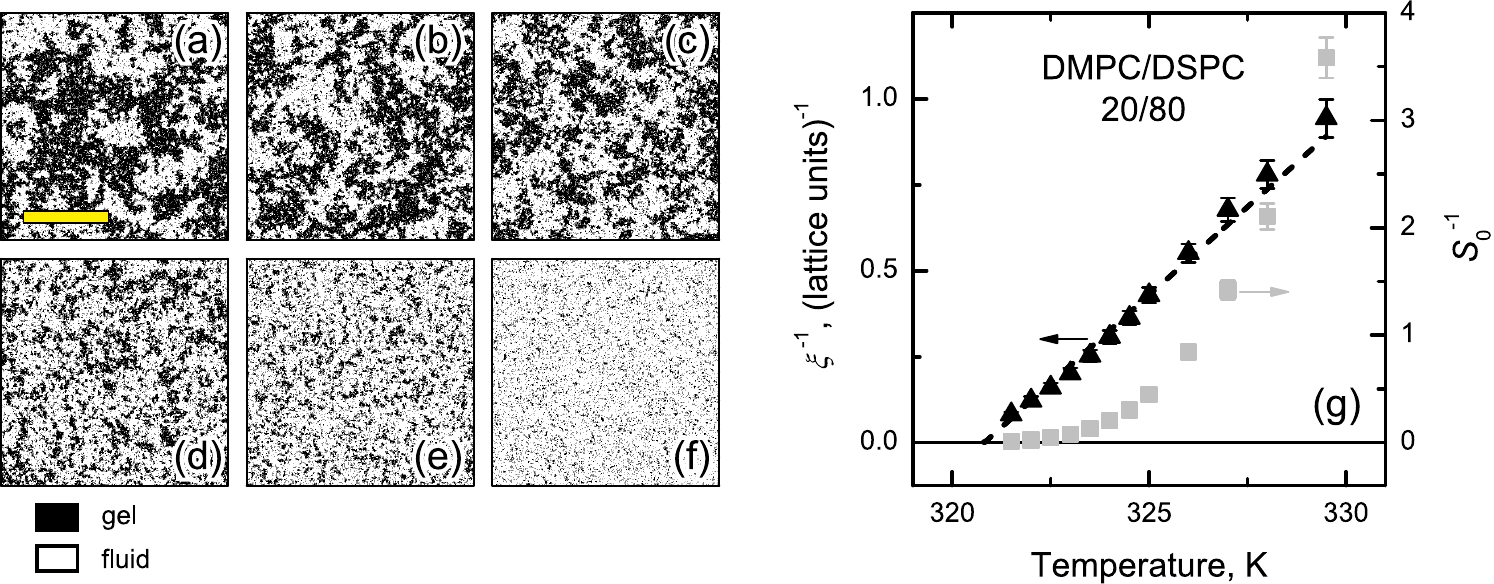}
\caption{(a-f) Representative equilibrium membrane configurations of the DMPC/DSPC 20/80 lipid mixture at $T = 320.7$ (a), 321.0 (b), 321.5 (c), 322.5 (d), 324.0 (e), and 328.0 K (f) obtained in MC simulations. The fluid and gel phases are shown with white and black, respectively. Lattice size: $400 \times 400$; scale bar: 150 lattice units $\approx 120$ nm. (g) Inverse correlation lengths (\fulltriangle) and inverse amplitude of the structure function (\fullsquare). A linear fit of the temperature dependence of the inverse correlation length (\dashed) gives an estimate of the critical temperature $T_{\rm c} = (320.8 \pm 0.1)$ K.}
\label{Fig_Xi_S0}
\end{center}
\end{figure}

For example, the correlation length of fluctuations and order parameter susceptibility in a supercritical system diverge upon approaching the critical point with exponents $\nu$ and $\gamma$, respectively \cite{Fisher1964}. One therefore should expect that the inverse correlation length and inverse amplitude of the structure function (proportional to the inverse order parameter susceptibility) should vanish close to the phase transition as
\begin{equation}
\xi^{-1}(T) \sim (T/T_{\rm c} - 1)^\nu
\label{eq:criticalXi}
\end{equation}
and
\begin{equation}
S_0^{-1}(T) \sim (T/T_{\rm c} - 1)^\gamma.
\label{eq:criticalS0}
\end{equation}
As it is exemplified in figure \ref{Fig_Xi_S0}, the inverse correlation length of fluid and gel domains for the DMPC/DSPC 20/80 mixture indeed tends to zero when the system is gradually cooled down to approach the two-phase fluid--gel coexistence region. A linear fit of this dependence yields an estimate of the critical temperature of $T_{\rm c} = (320.8 \pm 0.1)$ K.

In a similar manner, when approaching a critical point from the two-phase coexistence side, one should generally expect that the line tension vanishes according to the following law \cite{Widom1965}:
\begin{equation}
\lambda(T) \sim (1 - T/T_{\rm c})^\mu.
\label{eq:criticalLambda}
\end{equation}
Indeed, in the case of the DMPC/DSPC 20/80 mixture (figure \ref{Fig_LineTension}), which is very close to the critical composition, the line tension $\lambda$ vanishes as the critical temperature $T_{\rm c}$ is approached and can be well described by a linear dependence $\lambda(T) \sim (1 - T/T_{\rm c})$. The linear fit yields $T_{\rm c} = (320.5 \pm 0.4)$ K.

These two estimates of the critical temperatures are in good agreement with each other and reproduce well the position of the critical point estimated using the analysis depicted in figure \ref{Fig_CriticalPoint}.

\begin{figure}[!t]
\begin{center}
\includegraphics{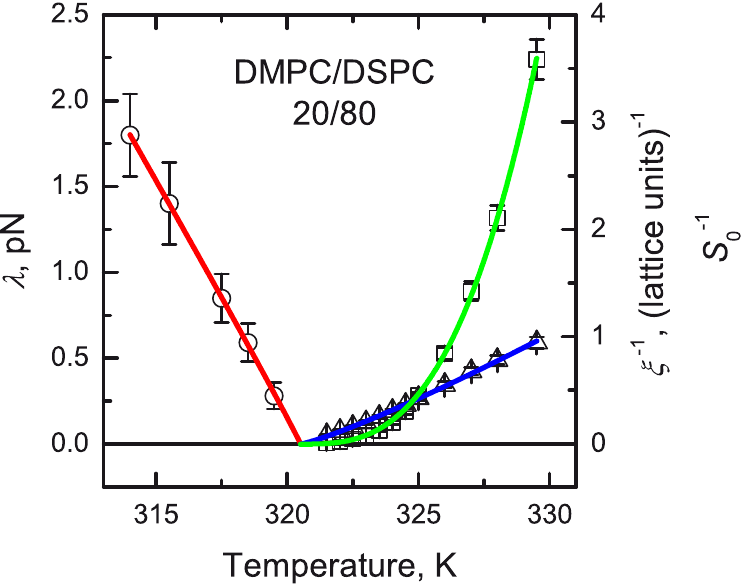}
\caption{Temperature dependences of the domain line tension $\lambda(T)$ (\opencircle) below the phase transition temperature, and of the inverse correlation length $\xi^{-1}(T)$ (\opentriangle) and inverse amplitude of the structure functions $S_0^{-1}(T)$ (\opensquare) above the phase transition obtained in MC simulations of the DMPC/DSPC 20/80 lipid system in the vicinity of the critical point. Curves show the global fit of these dependences to the expressions (\ref{eq:criticalXi}), (\ref{eq:criticalS0}) and (\ref{eq:criticalLambda}) with a common $T_{\rm c}$. The fit yields the estimates of the critical temperature $T_{\rm c} = (320.5 \pm 0.2)$ K and critical exponents $\mu = 1.17 \pm 0.04$, $\nu = 0.97 \pm 0.05$, and $\gamma = 2.94 \pm 0.05$.}
\label{Fig_LT_Xi_S0}
\end{center}
\end{figure}

We can attempt a more precise determination of the critical temperature, as well as the critical exponents of our system by simultaneously fitting the temperature dependences of the inverse correlation length, inverse amplitude of the structure function, and line tension in the vicinity of the critical point to expressions (\ref{eq:criticalXi}), (\ref{eq:criticalS0}) and (\ref{eq:criticalLambda}) with a common critical temperature $T_{\rm c}$. The results of this fit are shown in figure \ref{Fig_LT_Xi_S0}. The estimate of the critical temperature is again close to the above values: $T_{\rm c} = (320.5 \pm 0.2)$ K. The estimates of the critical exponents obtained in this fit are as follows: $\mu = 1.17 \pm 0.04$, $\nu = 0.97 \pm 0.05$, and $\gamma = 2.94 \pm 0.05$.

It should be noted that different physical systems often show the same set of critical exponents, and in this case are said to belong to the same universality class. This means that close to the phase transition particular microscopic details of a system become unimportant, and its behaviour is governed by a small number of features, such as dimensionality and symmetry. Recently, it was suggested that three-component membranes close to the critical point might belong to the 2D Ising universality class \cite{Honerkamp-Smith2008, Veatch2008}. For 2D systems in the Ising universality class the critical exponents take the following values \cite{Fisher1964, Hohenberg1977}: $\mu = 1$, $\nu = 1$, and $\gamma = 7/4$. While the fit in figure \ref{Fig_LT_Xi_S0} gives the estimates of the exponents $\mu$ and $\nu$ ($1.17 \pm 0.04$ and $0.97 \pm 0.05$, respectively) close to those expected for a 2D Ising model, the estimate of the exponent $\gamma$ ($2.94 \pm 0.05$) absolutely does not fit the predictions for the 2D Ising model.

We explain this strong discrepancy of the exponent $\gamma$ from the Ising model-based expectations by the fact that our system is different from the Ising model in one important aspect. Namely, while the Ising system above the phase transition is characterized by a zero average magnetization (on average equal numbers of up and down spins, or black and white pixels in a graphic representation), our membrane, having an equal amount of lipids in the fluid and gel conformations at the critical point (i.e., having a zero magnetization in terms of the Ising model), very fast transforms into the all-fluid state with increasing the temperature (i.e., in terms of the Ising model, its absolute average magnetization tends to 1). Our preliminary results show that, if, instead of keeping the membrane composition fixed, one would approach the critical point along the curve defined by the condition $X_{\rm fluid} = X_{\rm gel} = 0.5$ (cyan curve in figure \ref{Fig_CriticalPoint}), the exponent $\gamma$ becomes close to the value of 7/4 expected for the 2D Ising system.

Thus, based on this behaviour, we conclude that, generally, multicomponent membranes should not necessarily show the Ising criticality.

On the other hand, if the amounts of lipids in the fluid and gel states both stay close to 0.5 within a wide temperature range around a critical point, one can indeed expect to observe critical behaviour close to that of the Ising model. It would be interesting to check experimentally whether this condition holds for three-component model membranes and giant plasma membrane vesicles whose behaviour close to the critical point was recently claimed to follow the Ising criticality \cite{Honerkamp-Smith2008, Veatch2008}.

\subsection{Domain growth dynamics and dynamic scaling}

When the membrane, initially kept at a high temperature, where it is in the all-fluid state, is abruptly cooled down (quenched) to a temperature corresponding to fluid--gel phase coexistence, phase separation takes place, and domains of the gel and fluid phase start to appear and coarsen with time. Studying the kinetics of membrane domain coarsening after a sudden temperature quench into the phase coexistence region can reveal information on the mechanisms involved in phase separation \cite{Bray1994} and thus provide further understanding of the microscopic membrane organization and the interplay between the conformational (fluid--gel) state of lipids and their lateral diffusion.

Domain growth in lipid membranes upon a quench to the two-phase coexistence region has been addressed in several experimental works on three-component \cite{Garcia-Saez2007, Jensen2007, Saeki2006} and two-component \cite{Hu2006} lipid systems and MC simulation studies of two-component lipid systems \cite{Joergensen1995, Joergensen2000}. We are not aware of any publication where a systematic study of the dependence of the domain growth and dynamic scaling would be carried out for a specific lipid mixture as a function of its composition and/or quench temperature. In this section, we briefly address this issue based on the results of our MC simulations.

After the very early domain growth stage, during which small domains nucleate and become unstable, the stage with the power law growth of the mean size of domains $R(t) \sim t^n$ sets in, with the growth exponent $n$ being characteristic of the system type and growth mechanism \cite{Bray1994, Furukawa1985}. For systems with a non-conserved order parameter $n = 1/2$, whereas for systems with a conserved order parameter the growth exponent can take values of $n = 1/3$ or $n = 1/4$, depending on the particular mechanism of the domain growth.

In case where the growth is controlled by evaporation of smaller domains, and larger domains grow due to diffusive transport of material through the medium from domain boundaries with larger curvature to domain boundaries of smaller curvature (Ostwald ripening), one should expect the domain growth with the exponent $n = 1/3$, known as the Lifshitz--Slyozov--Wagner growth \cite{Lifshitz1961, Wagner1961}. Originally, this law was derived for the growth of minority phase domains in a three-dimensional system in the limit of small minority phase concentration. Later it was argued that the same law also applies to two-dimensional systems, and, based on results of simulations, it was suggested that it should also apply for any concentration of the minority phase \cite{Huse1986, Amar1988, Rogers1988, Rogers1989}. 

When spinodal decomposition takes place in a system with a conserved order parameter one has to distinguish two possible scenarios. While for (close to) symmetric quenches, i.e. (close to) equal amounts of the two phases, the spinodal decomposition will necessarily produce a worm-like bicontinuos morphology, in case of strongly asymmetric quenches this situation is generally not expected - the minority phase occupies a too low area to form a network and no bicontinuous structure, usually believed to be characteristic for spinodal decomposition, will be observed. The notion of spinodal decomposition in this case will just mean that the system is instable with regard to small concentration fluctuations \cite{Strobl1997}.

For both symmetric and asymmetric quenches, domains initially grow with $n = 1/4$ when spinodal decomposition takes place. For symmetric quenches this is the result of diffusion of particles along the interface boundaries of the bicontinuous phase morphology. If the quench is (strongly) asymmetric and thus the phase morphology after the quench is not bicontinuous, but rather appears as droplets of the minority phase embedded in the majority phase, the growth proceeds mainly via the Brownian motion and coalescence of these droplets for which in 2D $n = 1/4$ as well \cite{Furukawa1985}. For both cases, with coarsening of domains, the diffusion through the medium becomes progressively more important and the domain growth crosses over to the asymptotic regime with $n = 1/3$ \cite{vanGemmert2005}. Notice that in some cases the observation of the asymptotic growth with $n = 1/3$ may require extremely long simulations and large lattice sizes \cite{vanGemmert2005}.

\begin{figure}[!p]
\begin{center}
\includegraphics{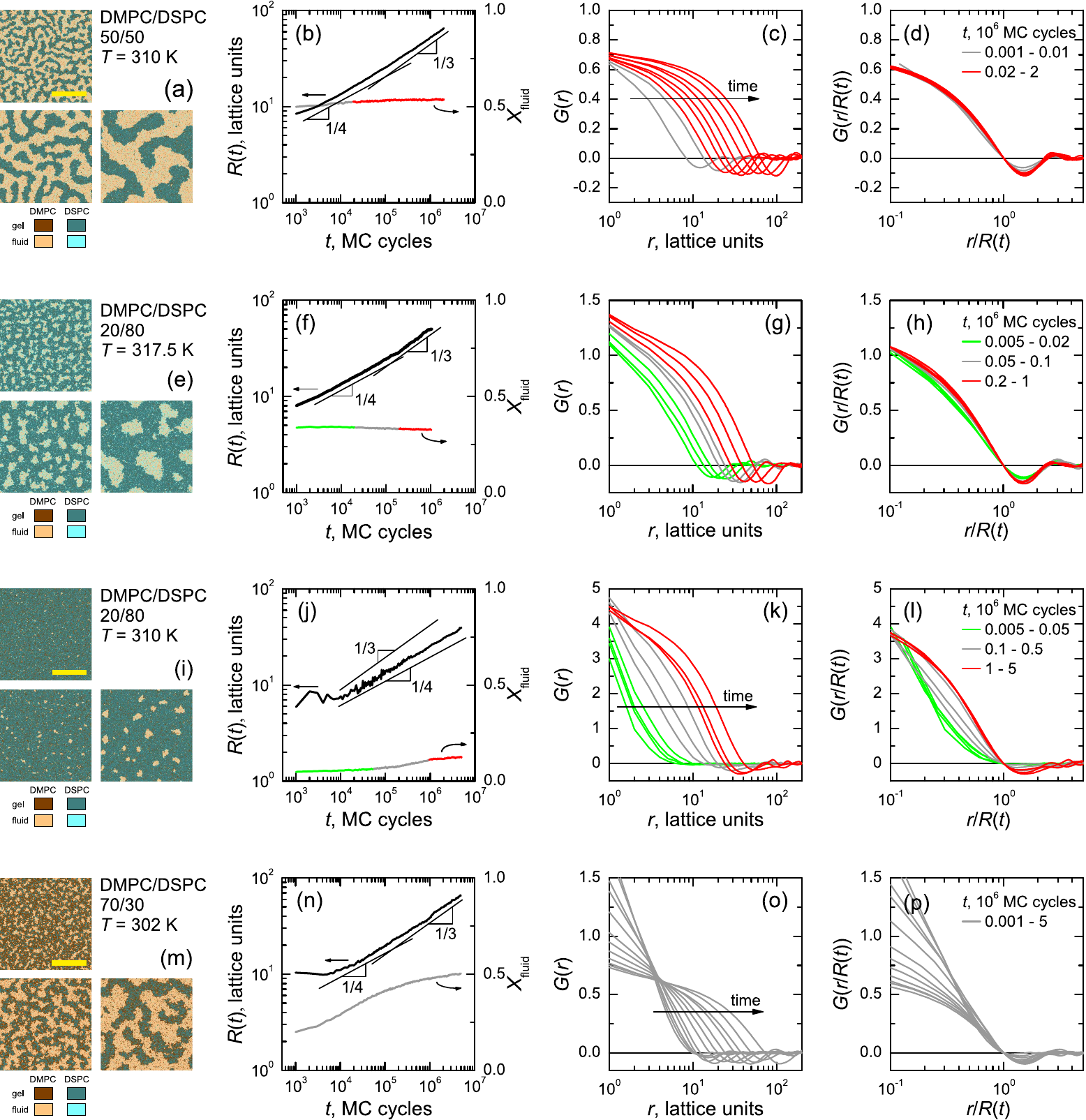}
\caption{Domain coarsening in DMPC/DSPC 50/50 (a-d), 20/80 (e-h), 20/80 (i-l) and 70/30 (m-p) membranes after a sudden temperature quench from $T = \infty$ to $T = 310$, 317.5, 310, and 302 K, respectively. Panels (a, e, i, m) each show three corresponding non-equilibrium configurations of the membrane at times $t = 10^4$ , $10^5$, and $10^6$ MC cycles after the quench. Lattice size: $400 \times 400$; scale bar: 150 lattice units $\approx 120$ nm. (b, f, j, n) Time dependence of the characteristic domain size $R(t)$ (black curves); thin black lines represent power laws with the exponent of 1/3 and 1/4. Also shown are the time dependences of the total fluid fraction $X_{\rm fluid}(t)$; the time intervals where $X_{\rm fluid}(t) \approx {\rm const}$ are marked by the green and red colours, while the time intervals where $X_{\rm fluid}(t) \ne {\rm const}$ are plotted in grey. (c, g, k, o) Radial autocorrelation functions $G(r)$  for a set of times (stated in panels (d, h, l, p)) after the quench. (d, h, l, p) Same as (c, g, k, o), respectively, replotted as a function of the reduced distance $r/R(t)$  to demonstrate that dynamic scaling is observed when $X_{\rm fluid}(t) \approx {\rm const}$; the colours of the curves at different time instants are chosen to correspond to those of $X_{\rm fluid}(t)$.}
\label{Fig_DynamicScaling}
\end{center}
\end{figure}

To study the growth of domains after an abrupt quench of the membrane being originally in the fluid state to a temperature within the coexistence region of the phase diagram, we calculated the radial autocorrelation function $G(r)$ (\ref{eq:G}) for a set of times $t$ after the quench and extracted the time-dependent domain size $R(t)$, which was defined as a distance $r$ at which the first zero-crossing of $G(r)$ occurs. As it is exemplified in figure \ref{Fig_DynamicScaling}, a power law growth with exponents ranging from $n = 1/4$ to $n = 1/3$ is observed in our simulations, in agreement with the theoretical expectations.

We should remark here again that our system is different from the ones typically discussed in the literature on coarsening dynamics, where usually either conserved order parameter or non-conserved order parameter systems are studied. Our system, however, cannot be strictly classified into either conserved- or non-conserved order parameter case. While the lipid composition in our model system is conserved, the fraction of phases is free to change. On the other hand, the fact that we observe a power law growth of domains with exponents ranging from $n = 1/4$ to $n = 1/3$, i.e. the ones characteristic for the conserved order parameter systems, suggests that the lipid demixing plays an important part in the phase separation process.

Remarkably, by comparing the domain growth shown in figure \ref{Fig_DynamicScaling}(b) with the experimental data on domain growth in double supported bilayers \cite{Jensen2007} we find that our simulations reproduce well the absolute rate of domain growth. In particular, extrapolation of the growth law to the later times shows that it takes of order of 100 s for the domains to grow to an average radius of  $\sim 1$ $\mu$m, in agreement with experiments \cite{Jensen2007}.

We also studied whether dynamic scaling \cite{Gunton1983, Henkel2010} is observed for domain coarsening in the membrane after a quench, i.e. whether the $G(r)$ corresponding to different time instants $t$ after the quench would collapse onto one master curve if replotted as a function of the reduced radius $r/R(t)$. It turned out that dynamic scaling is observed not in all cases, but only when the total fluid fraction $X_{\rm fluid}$ of the membrane is constant in time. The data presented in figure \ref{Fig_DynamicScaling} demonstrate this observation.

Figure \ref{Fig_DynamicScaling} (a-d) shows an example for a close to symmetric quench to a point of the phase diagram located deep within the region between the two spinodal curves: the membrane with DMPC/DSPC 50/50 was quenched to $T = 310$ K. The fluid fraction $X_{\rm fluid}$ remains constant during domain coarsening, except for a very short initial growth stage. At the same time, the observed growth exponent crosses over from $n \approx 1/4$ at early times to $n \approx 1/3$ at later times. This suggests that the growth first proceeds via diffusion along the interface ($n = 1/4$) and later via bulk diffusion from smaller evaporating domains to larger ones ($n = 1/3$). This behaviour can indeed be observed from the time series of snapshots from the simulation. Dynamic scaling of $G(r)$ is observed for the whole time interval where $X_{\rm fluid}(t) = {\rm const.}$

In figure \ref{Fig_DynamicScaling} (e-h) an example for a DMPC/DSPC 20/80 membrane that was quenched to $T = 317.5$ K is shown. Here, the amounts of the fluid and the gel phase are clearly not equal. Therefore, in this situation we observe an asymmetric quench of the membrane and the phase morphology, unlike in the former example, is not expected to be bicontinuous. The mean domain size initially grows with an exponent of $n \approx 1/4$, in agreement with the observed domain growth via the coalescence mechanism at early times, and later crosses over to $n \approx 1/3$, due to the domain growth via the evaporation of smaller domains at later times. The total fluid fraction in the membrane is nearly constant for up to $t \approx 2\times 10^4$ MC cycles and the $G(r)$ shows dynamic scaling in this time range. At later times $t \approx (0.5-1)\times 10^6$ MC cycles the fluid fraction changes -- most probably due to the rearrangement of lipids between the fluid and gel phases -- and no dynamic scaling is observed. Finally, at even later times $t > 2 \times 10^6$ MC cycles the fluid fraction becomes constant again, and dynamic scaling is again observed.

A very similar behaviour is observed for a quench of a DMPC/DSPC 20/80 membrane to $T = 310$ K (figure \ref{Fig_DynamicScaling} (i-l)): we observe two scaling regimes, both for time intervals where the fluid fraction of the membrane is constant, i.e. at times $t \lesssim 5\times 10^4$ MC cycles and $t \gtrsim 1\times 10^6$ MC cycles. No dynamic scaling is observed at the intermediate times when the fluid fraction of the membrane is changing. Notice that, unlike in the former two examples, the mean domain size grows as $R(t) \sim t^{1/4}$  for the whole time range that was studied.

Figure \ref{Fig_DynamicScaling} (m-p) shows the results for domain coarsening in a DMPC/DSPC 70/30 membrane quenched to $T = 302$ K. The mean domain size $R(t)$ shows a behaviour qualitatively similar to the first example of the DMPC/DSPC 50/50 membrane, quenched to $T = 310$ K, and grows with an exponent that crosses over from $n \approx 1/4$ at early times to $n \approx 1/3$ at later times, in agreement with the behaviour observed from a time series of membrane snapshots. Nevertheless, no dynamic scaling is observed in this case. Notice that the total fluid fraction under these conditions keeps changing over the whole time interval covered by the MC simulation. Remarkably, this change in the fluid fraction also changes the morphology of the phases. While the early stage is characterized by the appearance and growth of fluid droplets, at the later stage, when the fluid fraction grows to a value close to 0.5, the phase morphology changes to bicontinuous.

We conclude that this intriguing behaviour reflects a complex interplay between fluid--gel phase separation and lipid demixing. The equilibration of the total fluid fraction also needs lipid rearrangement via lateral diffusion of lipids, and the timescale of this process depends on the particular point in the phase diagram. This can serve as a potential explanation for the discrepancy in the experimental data on domain growth dynamics in lipid bilayers giving growth exponents from $n \approx 1/3$ \cite{Jensen2007, Garcia-Saez2007} down to $n \approx 0.15$ \cite{Saeki2006} (we should point out here that a careful analysis of the results published in \cite{Saeki2006} shows that that the late-stage growth kinetics exhibits the exponent close to 1/4, rather than 0.15 reported by the authors). The evolution of the total fluid fraction with time after the membrane quench to the two-phase coexistence region was previously observed in simulations \cite{Joergensen1995, Joergensen2000}, and it was suggested on a qualitative level that non-equilibrium effects may influence the dynamics of phase separation on various length and time scales. Our results clearly show that, indeed, the dynamic scaling can be strongly affected by these non-equilibrium phenomena.

In this context it is interesting to mention a recent simulation study \cite{Fan2010d} which showed that the presence of hydrodynamic interactions both within the membrane and in the surrounding solvent can lead to the breakdown of dynamic scaling upon spinodal decomposition in critical lipid mixtures. Our results thus point to an alternative mechanism which can lead to the breakdown of dynamic scaling in phase-separating membranes, which is related to the lateral redistribution of the membrane components. In experimentally realizable situations, we believe, both of the effects should be present and may enhance each other.

We believe that further insight into these issues can be gained by applying the theory of non-equilibrium phase transitions and ageing \cite{Henkel2010} and by comparing the outcome of the simulation results with outcomes of appropriately designed experiments.

\section{Conclusions}
In this paper, we have demonstrated that the properties of lipid membranes on the experimentally relevant spatial scales of order of a micrometer and time ranges of order of a second can be successfully addressed via lattice-based Monte Carlo simulations with very moderate computational efforts.

Using the DMPC/DSPC mixture as an example of a two-component lipid membrane, our large-scale simulations allowed us to obtain the following important results, which, to the best of our knowledge, were previously not reported in simulation studies of lipid membranes:

We found that within a certain range of lipid compositions, the phase transition from the fluid phase to the fluid--gel phase coexistence proceeds via near-critical fluctuations, while for other lipid compositions this phase transition has a quasi-abrupt character. Qualitatively, this is in excellent agreement with recent experiments where these two scenarios of the phase transition were observed in three-component lipid membranes \cite{Veatch2008}.

In the presence of near-critical fluctuations, close to the critical point of the membrane, lipids show transient subdiffusion, which is important for understanding the origins of anomalous diffusion in cell membranes, especially in view of the recent experimental results \cite{Veatch2008}, showing the critical behaviour in giant plasma membrane vesicles isolated from living cells.

In the fluid--gel phase coexistence region, macroscopic phase separation takes place, and after full equilibration of the membrane, a single circular-shaped domain of the minority phase emerges. Analysis of the thermally-induced domain shape fluctuations allowed us to obtain the line tension between the fluid and gel phases. We found that in the fluid--gel coexistence region, away from the critical point, the line tension is $\approx$ 2 pN, in good agreement with the available literature data on line tension in lipid membranes \cite{Honerkamp-Smith2008, Baumgart2003, semrau2009, Esposito2007, Baumgart2005,  Tian2007, Garcia-Saez2007, Blanchette2007}.

When approaching the critical point, the line tension, as well as the inverse correlation length of fluid--gel spatial fluctuations, and the corresponding inverse order parameter susceptibility of the membrane, approach zero. This is in agreement with recent experimental observations of the approach to criticality in three-component lipid bilayers and giant plasma membrane vesicles \cite{Honerkamp-Smith2008, Veatch2008}. On the other hand, contrary to the conclusions drawn in \cite{Honerkamp-Smith2008, Veatch2008}, we find that our system is not in the Ising universality class and provide a tentative explanation for this discrepancy.

An analysis of the domain coarsening dynamics after an abrupt quench of the membrane to the fluid--gel coexistence region revealed that the domain growth law may depend on the lipid composition and temperature, which reflects a complex interplay between fluid--gel phase separation and lipid demixing. We find that the domain growth exponent varies from 1/4 to 1/3, which is in a qualitative agreement with experimental data for lipid bilayers, for which a range of growth exponents has been reported as well \cite{Garcia-Saez2007, Jensen2007, Saeki2006}. We found out that the dynamic scaling of the radial autocorrelation function of the spatial distribution of the lipid state over the membrane during the domain coarsening is observed only when the fluid fraction of the membrane stays constant in time, which again shows the importance of lipid diffusion in the domain coarsening dynamics.

\section*{Acknowledgements}
The authors acknowledge inspiring discussions with Herv\'{e} Rigneault and Cyril Favard at the early stage of the project. Helpful comments by Richard L. C. Vink are gratefully appreciated.

The work was supported by the Deutsche Forschungsgemeinschaft via Research Group FOR 877 `From Local Constraints to Macroscopic Transport'.



\providecommand{\newblock}{}

\end{document}